\documentclass[a4paper,fleqn,usenatbib,useAMS]{mnras}
\usepackage[T1]{fontenc}
\usepackage{ae,aecompl}
\usepackage{graphicx}	
\usepackage{amsmath}	
\usepackage{amssymb}	

\title[GK Per during the 2015 outburst]{Multi-mission observations of the old nova GK Per during the 2015 outburst}

\author[P. Zemko, M. Orio, G. J. M. Luna, K. Mukai, P. A. Evans, A. Bianchini]{P. Zemko$^{1}$\thanks{E-mail:
polina.zemko@studenti.unipd.it},
M. Orio$^{2,3}$, 
G. J. M. Luna$^{4,5}$,
K. Mukai$^{6,7}$,
P. A. Evans$^{8}$,
and A. Bianchini$^{1,2}$
\\
$^{1}$Department of Physics and Astronomy, Universit\`a di Padova, vicolo dell' 
Osservatorio 3, I-35122 Padova, Italy\\
$^{2}$INAF - Osservatorio di Padova, vicolo dell' Osservatorio 5, I-35122 Padova, Italy\\
$^{3}$Department of Astronomy, University of Wisconsin, 475 N. Charter Str., 
Madison, WI 53704, USA\\
$^{4}$Universidad de Buenos Aires, Facultad de Ciencias Exactas y Naturales, Av. Inte. G\"uiraldes 2620, 
C1428ZAA, Buenos Aires, Argentina\\
$^{5}$CONICET-Universidad de Buenos Aires, Instituto de Astronom\'ia y F\'isica del Espacio, (IAFE), Av. Inte. G\"uiraldes 2620, 
C1428ZAA, Buenos Aires, Argentina\\
$^{6}$CRESST and X-ray Astrophysics Laboratory, NASA Goddard Space Flight Center, Greenbelt, MD 20771, USA\\
$^{7}$Department of Physics, University of Maryland, Baltimore County, 1000 Hilltop Circle, Baltimore, MD 21250, USA\\
$^{8}$Department of Physics and Astronomy, University of Leicester, Leicester LE1 7RH, UK\\
}

\date{}

\pubyear{2002}

\begin{document}

\label{firstpage}
\pagerange{\pageref{firstpage}--\pageref{lastpage}}
\maketitle
\begin{abstract}

GK Per, a classical nova of 1901, is thought to undergo variable mass accretion on to a 
magnetized white dwarf (WD) in an intermediate polar system (IP).
We organized a multi-mission observational campaign in the X-ray and ultraviolet (UV) energy
ranges during its dwarf nova (DN) outburst in 2015 March-April.
Comparing data from quiescence and near outburst, we have found that the maximum plasma 
temperature decreased from about 26 to 16.2$^{+0.5}_{-0.4}$ keV. This is consistent with the 
previously proposed scenario of 
increase in mass accretion rate while the inner radius of the magnetically disrupted 
accretion disc shrinks, thereby lowering the shock temperature.
A {\sl NuSTAR} observation also revealed a high-amplitude WD spin modulation of the very 
hard X-rays with a single-peaked profile, suggesting an obscuration of the 
lower accretion pole and an extended shock region on the WD surface. The X-ray spectrum 
of GK Per measured with the {\sl Swift} X-Ray Telescope 
varied on time-scales of days and also showed a gradual increase of the soft X-ray 
flux below 2 keV, accompanied by a decrease of the hard flux above 2 keV. 
In the {\sl Chandra} observation with the High Energy Transmission Gratings, we 
detected prominent emission lines, especially of Ne, Mg and Si, where the ratios of H-like
to He-like transition for each element indicate a much lower temperature than the 
underlying continuum. We suggest that the X-ray emission in the 0.8--2 keV 
range originates from the magnetospheric boundary.
 \end{abstract}


\section{Introduction}
GK Per underwent a nova explosion on 1901 February 21 \citep{wil1901gkper} and after a long period
of irregular fluctuations, in 1948, it started to behave like a dwarf nova (DN), with 
small amplitude (1 -- 3 mag.) outbursts lasting for up to two months and recurring at intervals
that can be roughly expressed as n$\times(400\pm40)$ days \citep{sab83gkper}. 
The widely accepted explanation of these outbursts is a recurring thermal instability in 
the inner part of the accretion disc (inside-out outbursts; see \citealt{bia86gkper}, \citealt{kim92tti}
 for the application of the disc instability model to GK Per). 

GK Per, at a distance of 470 pc \citep{mcl60gkperdist}, is the second closest nova
 to us and the best monitored at quiescence.
It is surrounded by an expanding nova shell, which emits in the X-rays 
\citep{bal05GKPerShell, tak15GKPershell},
weakly in UV \citep{eva92GKPerUV}, optical \citep[e.g.][]{sha12CKPer}, infrared 
\citep[e.g.][]{eyr99GKPerIR} and radio \citep{anu05GKPerradio} 
bands and has a complex structure with numerous knots \citep{lii12GKPerShellKin}
and a jet-like feature.

The orbital period is 1.9968 d \citep{cra86gkperorb, mor02gkper}. The secondary
is a K2-type subgiant mass of 0.25--0.48 M$_{\sun}$ \citep{war76gkper, wat85gkper, mor02gkper} 
and the mass of the primary is 0.86$\pm$0.02 M$_{\sun}$ \citep{sul16GKPer}. GK Per has an extended 
accretion disc ($\sim$10$^{11}$ cm) and interpreting the typical duration of
DN outbursts in GK Per as the viscous decay time-scale \citet{eva09GKPer} find that only about 10 \% of
the disc is involved in the outbursts.

\citet{kin79gkper} and \citet{bia83GKPerIP} proposed that GK Per hosts a magnetic
 white dwarf (WD). \citet{wat85gkper} discovered a 
modulation with a 351 s period in the X-ray light curve and proposed that this is 
the WD spin period. 
\citet{wat85gkper} also noticed that the pulse fraction is 50 \% above 3 keV and up to 80 \%
below 3 keV, but does not depend on the mean X-ray flux. The pulse profile is single-peaked
in outburst and double-peaked, with smaller amplitude in quiescence \citep{nor88GKPer, hel04GKPerRXTE}.

Apart from the spin-related modulation, during outbursts GK Per shows quasi periodic oscillations (QPOs) in X-rays
and in the optical band with the characteristic time-scale of several kiloseconds 
\citep[e.g][]{wat85gkper, mor99gkperqpo,nog02GKPer,hel04GKPerRXTE}.
This time-scale is exceptionally long compared to what is typically observed in CVs \citep{war04QPO}.
QPOs with periods of 360--390 s were also detected in optical observations of GK Per 
by several authors. 
\citet{wat85gkper} noticed that the beat period between the spin period and the
shorter-period QPOs is in the kiloseconds range, close to the period of the longer QPOs. \citet{mor99gkperqpo} 
explored this idea, assuming that the QPOs are due to absorption of the emission from
the innermost regions of the accretion disc by blobs of material in the magnetosphere.
The 360--380 s period in this model is due to the the Keplerian velocity in the inner disc radius.
\citet{hel04GKPerRXTE} argued that this scenario
 explains the optical data of \citet{mor99gkperqpo}, but does not fit 
the X-ray observations. These authors suggested instead that bulges of material
in the inner edge of the accretion disk, high above the disc plane, obscure the
WD and the  $\sim$5000s period is due to the slow-moving prograde
waves in the innermost region of the disc \citep{war02DNO}.

Several pieces of evidence indicate that during DN outbursts the accretion
disk in GK Per pushes the magnetosphere towards the surface of the WD \citep{hel04GKPerRXTE, 
vie05GKPer, sul16GKPer} as expected by theory \citep[see, e.g.,][]{gho79disk}. 
This scenario, first proposed by \citet{hel04GKPerRXTE} in analogy with XY Ari \citep{hel97XYAri}, 
explains the single-peaked pulse 
profile as a consequence of cutting off the line of sight to the lower accreting pole.
The maximum plasma temperature in GK Per in quiescence is higher than in outburst 
\citep{bru09IPBAT}, which can also be understood in the context of the shrinking of the 
accretion disc \citep{sul05IPmass}. However,
while the hard X-ray emission can be explained as due to the accretion columns emitting 
bremsstrahlung radiation that is highly absorbed, the broad-band X-ray spectrum of GK Per 
and its evolution during outbursts are much more complex.
\citet{sim15GKPerSatur} noticed that the halt of the rise of the hard X-ray luminosity 
is not modelled only by just assuming increased absorption, and probably involves large 
structural changes of the accreting regions. 
\citet{muk03twotypes} have shown that the outburst X-ray spectrum of GK Per below 2 keV is not 
consistent with a ``typical'' cooling-flow spectrum of an accreting WD and proposed a 
considerable contribution of photoionization. \citet{vie05GKPer} and \citet{eva07softIP}
noticed that the soft X-ray emission of GK Per detected during the 2002 outburst may originate
in the heated surface of the WD, like in the case of ``soft intermediate polars''.
 
On 2015 March 6.84 Dubovsky (VSNET-ALERT 18388) and Schmeer (VSNET-ALERT 18389) discovered that 
GK Per has started a new DN outburst and was at a magnitude 12.8. We proposed 
a multimission observation campaign in order to follow the evolution of the object during the 
outburst and to obtain X-ray spectra in a broad energy range, revealing the physical processes 
that take place in this binary system. 

\section{X-ray observations and data analysis}

We started the observations of the 2015 DN outburst of GK Per as soon as it became observable
with {\sl Swift} on March 12 2015, and we could follow the outburst almost until the optical maximum. 
We wanted a higher cadence than in the previous 2006 outburst {\sl Swift} monitoring,
and obtained two exposures per day for two weeks and one exposure per day in the following 
two weeks. During the first two weeks of the monitoring the {\sl Swift} X-Ray Telescope 
(XRT) \citep{bur05XRT} was in the automatic state \citep{hil05XRTauto}, choosing the photon
counting (PC) or window timing (WT) operating mode depending on the source count rate. 
During the second two weeks the monitoring was performed only in the WT mode.
The {\sl Swift} Ultraviolet Optical Telescope (UVOT) was in the image mode, providing the 
mean magnitude per exposure in one of the four UVOT filters (U, UVW1, UVM2, UVW2).
 Coordinated {\sl NuSTAR} and {\sl Chandra} Advanced Imaging Spectrometer High-Energy
Transmission Grating (ACIS-S/HETG) observations were performed on April 4 2015, close to the
optical maximum. The {\sl Chandra} ACIS-S/HETG consists of two sets 
of gratings \citep{can05HETG}, operating in the 31--2.5 {\AA} (Medium Energy Grating) and 
15--1.2 {\AA} (High Energy Grating) energy ranges with the resolution better than 0.023 {\AA}.
{\sl NuSTAR} has two detectors: FPMA and FPMB, covering the same energy range, from 3 to 79 keV.
The time resolution of all the instruments used for the observations is better than 2.5 s,
which is much shorter than the periodicities we expected to find.
The {\sl Swift} XRT spectra and the mean X-ray 
count rate per snapshot in the 0.3--10 keV energy range were obtained using the {\sl Swift} XRT data products generator 
\citep{eva09XRTsp, eva07XRTlc}, subsequently performing the barycentric correction with
an applet\footnote{
http://astroutils.astronomy.ohio-state.edu/}, created by \citet{eas10barycor}. 
The {\sl Swift} XRT light curves in the WT mode and in the PC mode (the latter 
excluding the central region and subsequently running \texttt{xrtlccorr}, because of the pile-up) and 
the UVOT data were processed with the \texttt{ftools} package. 
The {\sl Swift} XRT light curves were barycentric corrected using \texttt{barycorr} tool.


We also used the processed {\sl Swift} Burst Allert Telescope
(BAT) data from the {\sl Swift} BAT transient monitor page \citep{kri13swiftbat}.
The {\sl Chandra} data were reduced with \texttt{CIAO v.4.7} and the {\sl NuSTAR} data with
the standard \texttt{nuproducts} pipeline. The {\sl NuSTAR} and {\sl Chandra} light 
curves were barycentric corrected. 
The X-ray spectra were analysed and fitted with
\texttt{XSPEC} v. 12.8.2. The list of the observations with the exposure times and count 
rates is presented in table \ref{tab:obs}. 

\begin{table}
\begin{minipage}{80mm}
\caption{Observational log.}
\begin{tabular}{lccc}
\hline
\hline
Instrument 	   & Date$^*$   & Exp.(s)& Count rate	\\
\hline	
Swift XRT(PC)  &   57093.15 & 606.7  & 1.72$\pm0.10$ \\
		       &   57093.58 & 1268.7 & 2.10$\pm0.08$ \\
		       &   57094.08 & 1033.0 & 1.70$\pm0.08$ \\
          	   &   57094.68 & 373.6  & 1.15$\pm0.10$ \\
    	       &   57095.04 & 952.8  & 2.12$\pm0.13$ \\
    	       &   57095.93 & 960.3  & 1.41$\pm0.07$ \\
    	 	   &   57096.30 & 744.7  & 1.49$\pm0.09$ \\
               &   57096.90 & 441.3  & 1.79$\pm0.14$ \\
               &   57097.04 & 970.3  & 1.33$\pm0.08$ \\
               &   57097.57 & 937.7  & 0.94$\pm0.06$ \\
               &   57098.03 & 997.9  & 1.62$\pm0.09$ \\
               &   57098.97 & 882.6  & 1.35$\pm0.07$ \\
               &   57099.17 & 875.0  & 1.28$\pm0.07$ \\
               &   57099.57 & 514.0  & 1.78$\pm0.12$ \\
               &   57100.11 & 922.7  & 1.58$\pm0.08$ \\
               &   57100.97 & 1033.0 & 1.08$\pm0.06$ \\
               &   57101.10 & 1133.3 & 1.11$\pm0.06$ \\
               &   57101.57 & 925.2  & 1.64$\pm0.09$ \\
               &   57102.23 & 1058.0 & 1.18$\pm0.06$ \\
               &   57102.86 & 910.1  & 1.61$\pm0.09$ \\
               &   57103.10 & 917.7  & 0.85$\pm0.06$ \\
               &   57103.63 & 418.7  & 0.86$\pm0.07$ \\
               &   57104.22 & 140.4  & 1.07$\pm0.12$ \\
               &   57104.70 & 975.3  & 1.70$\pm0.10$ \\ 
               &   57105.95 & 905.1  & 0.67$\pm0.04$ \\
               &   57106.15 & 1075.6 & 1.72$\pm0.09$ \\
               &   57106.76 & 965.3  & 1.15$\pm0.07$ \\
               &   57107.29 & 1110.7 & 1.71$\pm0.09$ \\
               &   57107.82 & 1196.0 & 0.70$\pm0.04$ \\
\hline	
Swift XRT(WT)  &   57094.65 & 497.7  & 3.01$\pm0.08$ \\
 			   &   57106.15 & 28.7	 &  3.6$\pm0.5$ \\
 			   &   57108.75 & 985.2  & 1.24$\pm0.04$ \\
 			   &   57109.62 & 986.3  & 1.43$\pm0.04$ \\
 			   &   57110.51 & 1145.6 & 1.31$\pm0.04$ \\
  		       &   57111.28 & 1086.9 & 1.93$\pm0.04$ \\
   		       &   57112.85	& 1877.8 & 1.69$\pm0.03$ \\
		       &   57113.34 & 1228.2 & 1.84$\pm0.04$ \\
     	       &   57114.13 & 1040.9 & 1.18$\pm0.04$ \\
               &   57115.90 & 2401.5 & 1.35$\pm0.03$ \\
               &   57116.11 & 10801.0 & 1.46$\pm0.04$ \\
               &   57117.66 & 974.1  & 1.52$\pm0.04$ \\
               &   57118.20 & 903.6  & 2.18$\pm0.05$ \\
               &   57119.29 & 1261.4 & 1.45$\pm0.06$ \\
      	       &   57120.36 & 1578.7 & 1.26$\pm0.04$ \\
           	   &   57121.06 & 1049.5 & 1.17$\pm0.04$ \\           	   
\hline	
Chandra	MEG	   &   57116.83 & 69008	& 0.0751$\pm$0.0010\\
Chandra	HEG	   &   57116.83	& 69008 & 0.1214$\pm$0.0013 \\
\hline	
NuSTAR	FPMA   &   57116.12 & 42340 &  3.665$\pm$0.009  \\
NuSTAR	FPMB   &   57116.12 & 42340 &  3.626$\pm$0.009 \\
\hline	
\multicolumn{4}{p{.9\textwidth}}{{\bf Notes.}$^*$ Modified Julian Date. The count rates
were measured in the following energy ranges: {\sl Swift} XRT --- 0.3--10.0 keV, 
{\sl Chandra} MEG --- 0.4--5.0 keV, {\sl Chandra} HEG --- 0.8--10.0 keV, {\sl NuSTAR} FPM 
--- 3--79 keV.}\\
\hline
\end{tabular}
\label{tab:obs}
\end{minipage} 
\end{table}

\section{Development of the outburst}
\begin{figure*}
\centering
\includegraphics[width=400pt]{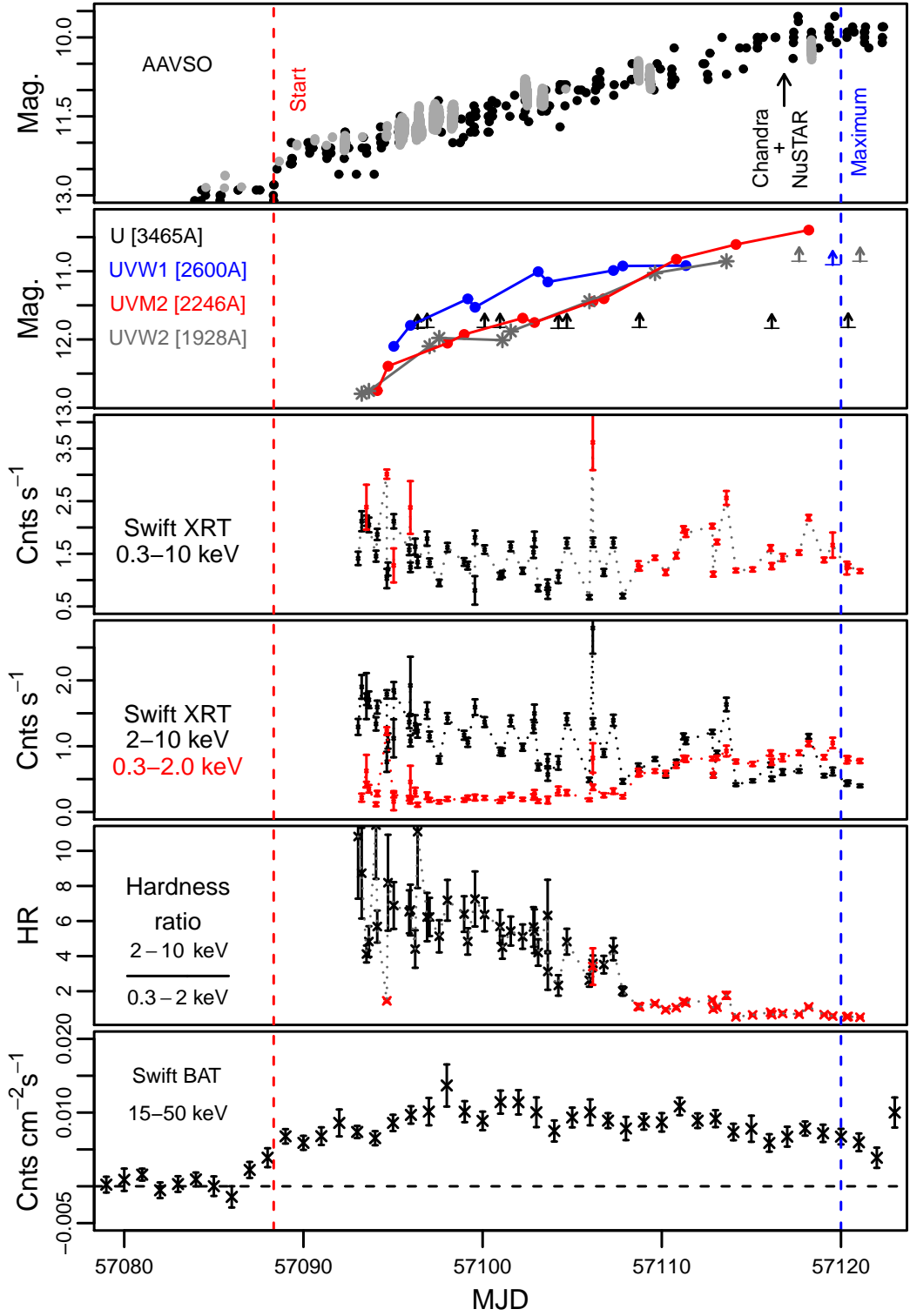}
\caption{From top to bottom: AAVSO light curve in the {\it V} band (grey) and without a 
filter (black). The red and blue vertical lines in all the panels mark the beginning
(MJD 57088.34) and the maximum (MJD 570120) of the outburst in the optical band. 
The {\sl Swift} UVOT light curves in different filters. 
The {\sl Swift} XRT light curve in the PC (red) and WT (black) modes. 
The {\sl Swift} XRT light curves above 2 keV (black) and below 2 keV (red).
The X-ray hardness ratio from the data obtained in the WT (red) and PC (black) modes. The {\sl Swift} BAT 
light curve.}
\label{fig:lc}
\end{figure*}

Figure \ref{fig:lc} shows a comparison between the development of the outburst in the optical,
UV and X-rays. The top panel is the optical light curve, which was obtained from
The American Association of Variable Star Observers (AAVSO)\footnote{https://www.aavso.org/}. 
The beginning of the outburst in the optical band was taken as a reference and is marked 
with the red dashed vertical line in all the panels. The maximum of the outburst 
in the optical band is also marked with the blue line. The second panel shows the {\sl Swift} 
UVOT data in different filters. All the UVOT light curves showed a gradual rise. 
The images in the lower energy filter --- U --- and the last observations in the 
 UVW1 and UVW2 filters were saturated and provide only lower limits for the magnitudes. 
We analysed also quiescence {\sl Swift} UVOT observations of GK Per obtained in 2012
to find the amplitude of the outburst in the UV range. Table \ref{tab:uvot} shows the 
quiescent UVOT magnitudes.

\begin{table}
\begin{minipage}{70mm}
\caption{The {\sl Swift} UVOT magnitudes of GK Per obtained in quiescence.}
\begin{tabular}{lccc}
\hline
\hline
MJD 	  & Filter & Exp. (s) & Mag. \\
\hline	
56115.284 & UVW1 & 361 & 15.69$\pm$0.02 \\
56115.279 & UVM2 & 413 & 16.75$\pm$0.03 \\
56115.274 & UVW2 & 413 & 16.67$\pm$0.02 \\
\hline	
56117.086 & UVW1 & 319 & 15.85$\pm$0.02 \\
56117.082 & UVM2 & 316 & 16.95$\pm$0.04 \\
56117.078 & UVW2 & 316 & 16.92$\pm$0.03 \\
\hline	
56118.014 & UVW1 & 143 & 15.67$\pm$0.03 \\
56118.013 & UVM2 & 112 & 16.74$\pm$0.05 \\
56118.012 & UVW2 & 112 & 16.68$\pm$0.04 \\
\hline
\end{tabular}
\label{tab:uvot}
\end{minipage} 
\end{table}

The third panel is the {\sl Swift} XRT light curve averaged within a snapshot in the whole energy range:
from 0.3 to 10 keV. The count rate varied from 0.7 to about 3
cnts s$^{-1}$ but did not show any significant increasing or decreasing trend.  
 
We missed the initial steep rise, observed only with the {\sl Swift} BAT,
 because GK Per was too close to the Sun, and observed only a plateau in the {\sl Swift} 
 XRT light curve.  In fig. \ref{fig:lc}, the bottom panels show the soft (0.3--2.0 keV)
and hard (2.0--10 keV) XRT light curves. The hard count rate is more scattered in comparison 
with the soft one and decreased as the outburst developed. The soft count rate, in turn, 
showed a prominent rise, which resulted in a gradual decrease of the hardness ratio
(panel 5) with minimum around MJD 57113, 25 days after the beginning of the outburst. 
The light curve measured with the 
{\sl Swift} BAT is more stable and only showed a moderate decrease after maximum around day
10 after the outburst. The flux increase started 2 days earlier in 
the {\sl Swift} BAT energy range than in optical.

\begin{figure*}
\centering
\includegraphics[width=400pt]{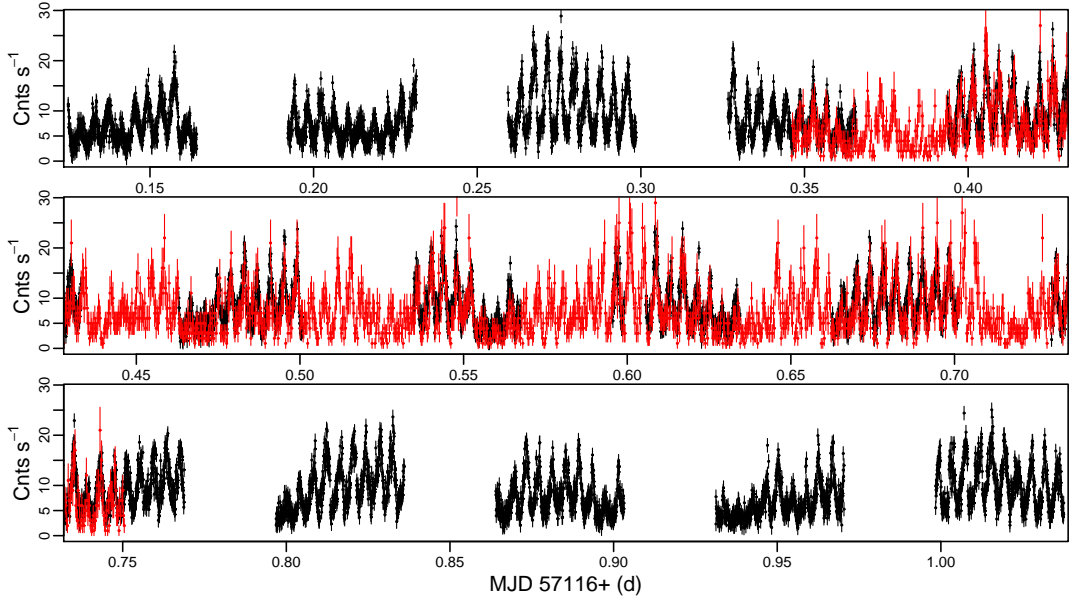}
\caption{{\sl NuSTAR} FPMA+FPMB light curve binned every 10 s (black) and the {\sl Chandra} HETG
light curve in the 1--6 {\AA} wavelength range, binned every 20 s (red) 
and multiplied by a factor of 20 for visibility.}
\label{fig:nuchan_lc}
\end{figure*}

\begin{figure*}
\centering
\includegraphics[width=340pt]{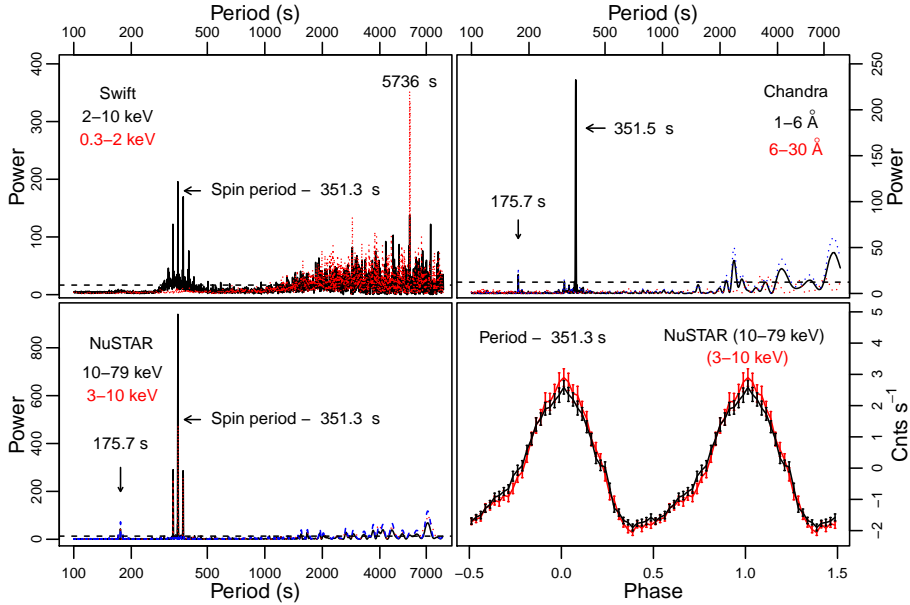}
\caption{Top-left: the LSP of the {\sl Swift} XRT data in the 2--10 keV energy
range (black) and at 0.3 -- 2 keV (red). Top-right: the LSP of the {\sl Chandra} HETG data in the
energy range 1--6 {\AA} before (red) and after (blue) subtracting the highest peak, at 351.5 s.
The red line shows the LSP of the {\sl Chandra} HETG data in the energy range 6--30 {\AA}.
Bottom-left: the LSP of the {\sl NuSTAR} data in the 10 -- 79 keV (black) and 3 -- 10 keV (red) energy
range. The blue line shows the LSP of the {\sl NuSTAR} data in the 10 -- 79 keV range 
after subtracting the peak at 351.3 s. 
The horizontal dashed lines show the 0.3\% false alarm probability level at all the LSPs.
Bottom-right: the {\sl NuSTAR} light curve in the  
3 -- 10 keV (red) and 10 -- 79 keV (black) ranges, folded with the WD spin period of 351.3 s}
\label{fig:timing}
\end{figure*}

\begin{figure*}
\centering
\includegraphics[width=350pt]{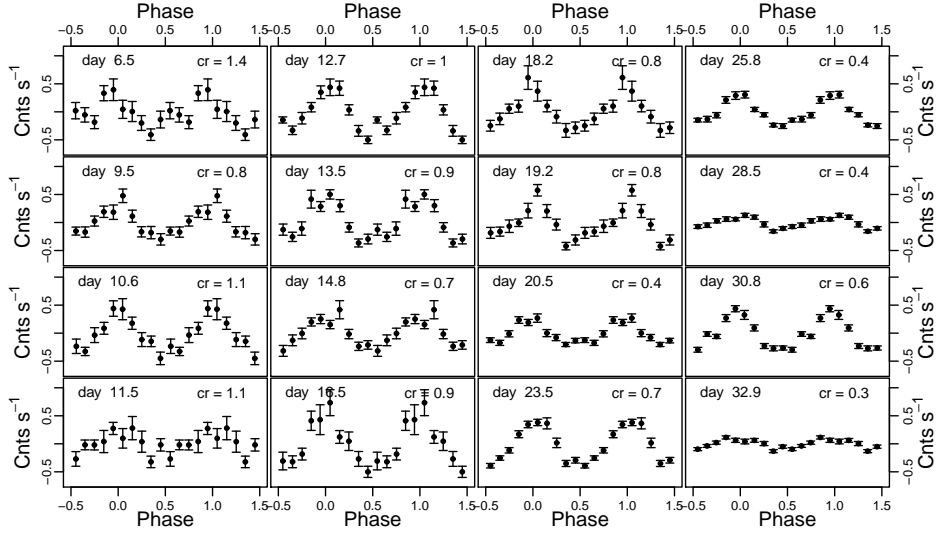}
\caption{Evolution of the spin pulse profiles in the {\sl Swift} XRT light curves above 2 keV.
We combined every three individual observations and folded them with the 351.3 s period. 
The mean date of observation and the mean count rate (cr) is marked on each plot.}
\label{fig:prof}
\end{figure*}

\begin{figure}
\centering
\includegraphics[width=250pt]{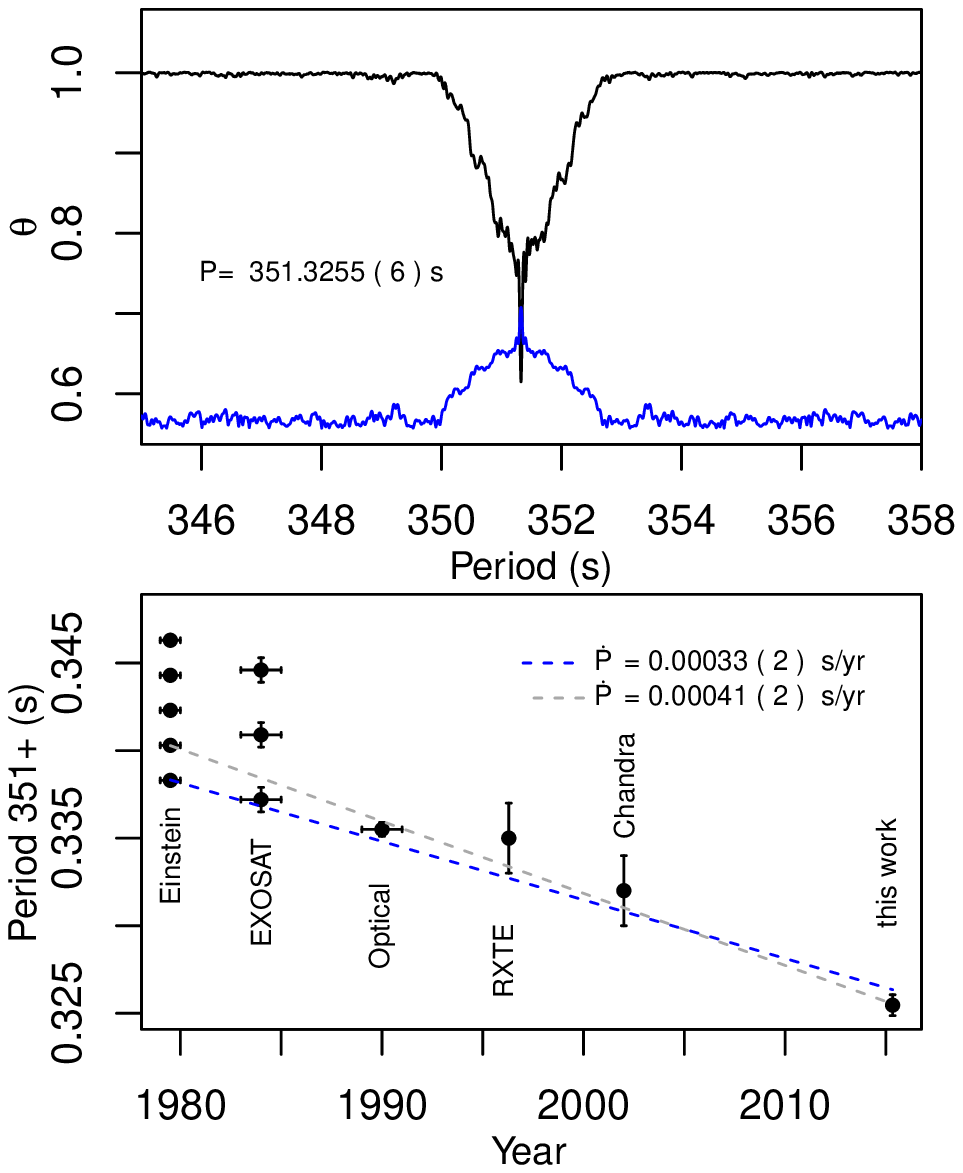}
\caption{Top: PDM analysis of the {\sl NuSTAR}, {\sl Chandra} HETG 1--6 {\AA} and hard {\sl Swift} XRT
light curves. Bottom: the WD spin period as a function of time and the
result of the linear fitting.}
\label{fig:spin}
\end{figure}

\section{Timing analysis}\label{timing}
\subsection{Broad band power spectra.}
For our timing analysis we extracted the {\sl Swift} XRT, {\sl NuSTAR} (the FPMA and FPMB 
detectors were combined) and {\sl Chandra} light curves and binned them every 10 s. 
In order to investigate a possible energy dependence of the X-ray variability we split the 
{\sl Swift} XRT light curve in two ranges: 0.3--2.0 and 2.0--10 keV, the {\sl Chandra} HETG 
light curve was extracted below 6 {\AA} and above 6 {\AA} and the {\sl NuSTAR} one above 
and below 10 keV.
{\sl Chandra} hard and {\sl NuSTAR} data show a strong periodic modulation, which can be seen 
in fig. \ref{fig:nuchan_lc}. For visibility in fig. \ref{fig:nuchan_lc} the {\sl Chandra}
count rate was multiplied by a factor of 20. To study this modulation quantitatively, we 
constructed Lomb-Scargle 
periodograms \citep{sca82LS} of various data in soft and hard energy bands.
The Lomb-Scargle periodograms (LSPs) in a broad range of periods for all the light 
curves are presented in fig. \ref{fig:timing}. The top-left-hand panel shows the LSPs 
of hard (black) and soft (red) {\sl Swift} XRT light curves.
The highest peak of the hard LSP corresponds to the WD spin period --- 351.33 s. 
The spin modulation is not present in the LSP of the soft {\sl Swift} XRT light curve:
it shows peaks only at longer time scales with the strongest one at 5736 s. 
Although QPOs in GK Per on a time-scale of $\sim$5000 were reported by many authors, this 
period is too close to the {\sl Swift} orbital period of 5754 s to be distinguished from the
windowing in the observations. The top-right- and bottom-left-hand panels of fig. \ref{fig:timing} show the
LSPs of the {\sl Chandra} HETG and the {\sl NuSTAR} light curves, respectively. There is no peak
in the Chandra soft LSP at 5736 s period, which confirms that this peak in the soft 
{\sl Swift} periodogram does not correspond to a real modulation. On the other hand, the
{\sl NuSTAR} and the hard {\sl Chandra} LSPs show peaks at $\sim$7000 s. These light curves are
indeed variable on the time-scales of kiloseconds, as 
shown by fig. \ref{fig:nuchan_lc}.

The blue dashed lines in the top-right- and bottom-left-hand panels show the LSPs after the subtraction
of the peak corresponding to the spin period. We fitted and subtracted a sine function with a 
fixed period of 351.33 s from the original data, and plotted the LSPs
again. The peak close to half of the spin period remained in the hard {\sl NuSTAR} and
{\sl Chandra} LSPs.

\subsection{Energy dependence of the WD spin modulation.}
All the LSPs of the light curves extracted above 2 keV show a prominent peak corresponding 
to the WD spin period, while neither {\sl Chandra} nor {\sl Swift} soft LSPs show any. 
The absence of the spin modulation in the region of 0.3 -- 2 keV may indicate that this 
emission component has a different origin and is visible during the whole spin cycle. 
We also extracted the light curves from the {\sl Chandra} HETG data in the regions of the strongest 
emission lines of Mg, Si and Fe K$\alpha$ 6.4 keV and checked whether the flux in these lines
is modulated with the orbital or with the spin period. Only the Fe K$\alpha$ 
line emission showed spin modulation, while the flux in the other
emission lines had aperiodic fluctuations. 

The peak corresponding to the spin period is present even in the LSP of the
{\sl NuSTAR} light curve above 10 keV. Typically the spin 
modulation of IPs is not detected, or only marginally measurable, in the hard X-rays
\citep{nor89Spin, muk15refl},
since the cross-section of the photoelectric absorption that usually causes the modulation 
decreases with energy. 
The effect of photoelectric absorption is not significant above 10 keV, so the observed 
high energy modulation might originate in a different mechanism than absorption of the 
accretion column emission. 
The comparison of the phase folded {\sl NuSTAR} light curves in two energy ranges (fig. \ref{fig:timing})
confirms that the spin modulation is not energy dependent: the spin profiles 
are almost identical. 

\subsection{Evolution of the spin pulse profile with the {\sl Swift} XRT.}
The next step was to explore the spin modulation of the light curves. We first investigated how 
the spin pulse profile changed as the outburst developed. Since the modulation is mostly seen
in hard X-rays, we combined the {\sl Swift} XRT light curves above 2 keV in groups
of three exposures and folded them with the 351.3 s period using a constant number of bins.
The result is shown in fig. 
\ref{fig:prof}: the pulse profile becomes smoother with time and stabilized 
by day $\sim$20 after the beginning of the outburst in optical. 
The pulse amplitude also depends on the mean value of the count rate. This will be further explored 
using the {\sl NuSTAR} data.

\subsection{WD spin-up rate.}
We measured the spin period more precisely combining the {\sl NuSTAR}, {\sl Swift} XRT 
light curve extracted above 2 keV and the {\sl Chandra} HETG light curve 
in the  1--6 {\AA}  wavelength range. Since the spin amplitude in each instrument is a similar 
fraction of the mean count rate, and the mean count rate is much lower for {\sl Swift} XRT 
and the {\sl Chandra} HETG, we first normalized the light curve from each instrument to its 
mean count rate. In order to account for 
possible long-term variability we removed linear trends from each segment the {\sl NuSTAR} 
light curve, from the {\sl Chandra} light curve and from each {\sl Swift} observation before 
applying the phase dispersion minimization (PDM) method \citep{PDM}. Fig. \ref{fig:spin} shows 
that the PDM analysis resulted in a value of the spin period P$_{\rm spin}$= 351.3255(6) s. 
Combining the WD spin measurements performed by 
\citet{era91GKPer,wat85gkper,nor88GKPer, pat91GKPer, hel04GKPerRXTE, mau04gkper} and following 
a discussion by \citet{pat91GKPer} and \citet{mau04gkper}, we fitted the trend
of P$_{\rm spin}$ as a function of time with a linear function. The uncertainty in the spin-up
rate is due to the uncertainty in the period derived from the {\sl Einstein} data:
two different values, 351.3383 and 351.3403 s, resulted in an acceptable fit, and adopting 
each of them yields a spin-up rate of 0.00033(2) s yr$^{-1}$ and 0.00041(2) s yr$^{-1}$,
respectively.

Using the new value of the spin period (P$_{\rm spin}$= 351.3255(6) s), we also calculated
the ephemeris of the pulse maxima. The Modified Barycentric Julian Date (MBJD) of the maxima can be
found as:
\begin{equation}
{\rm T_{\rm max}(MBJD)}=57093.51666(21)+0.004066267(7)*E
\end{equation}

\begin{figure}
\centering
\includegraphics[width=220pt]{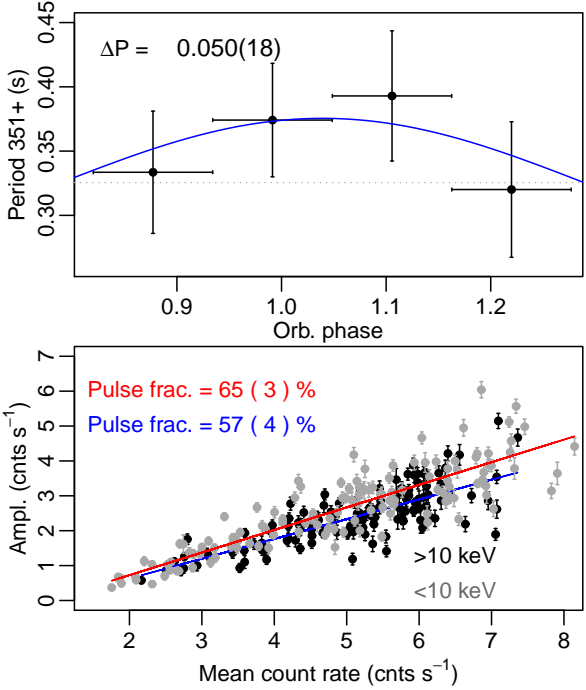}
\caption{Top: the WD spin period as a function of orbital phase and the
result of the fit with a sine function. The dashed grey line shows the mean period, 351.3255 s.
Bottom: the amplitude of pulses, calculated as (maximum-minimum)/2, as a function of 
the mean count rate per pulse and the result of the linear fit.
The grey dots and red line correspond to the {\sl Nustar} light curve, extracted below 10 keV,
and the black dots and blue line --- above 10 keV. }
\label{fig:oc}
\end{figure}

\subsection{Orbital variability of the spin period.}
\citet{wat85gkper} pointed out that the spin period should be modulated with the orbital 
one (i.e. X-ray radial velocity curve), but taking into account the estimates of the binary 
parameters of GK Per, the authors concluded that these variations would be below the detection 
limit. With the most recent measurements 
of the mass ratio $q$, WD mass $M_{\rm WD}$, binary inclination angle $i$ and orbital
period $P_{\rm orb}$, we derived the expected amplitude of variation of the pulse 
period $\Delta P_{\rm spin}$ with orbital phase.
 The semi-major axis of the WD's orbit is: 
\begin{equation}
a_x \sin{i}=\frac{P_{\rm orb}}{2 \pi} K_x
\end{equation}
where $K_x$ is the radial velocity semi-amplitude. Assuming a circular orbit \citep{cra86gkperorb} we can write:

\begin{equation}
M_2  \sin^3{i}=\frac{P_{\rm orb} K_x^3}{2 \pi G}\left(1+\frac{1}{q}\right)^2
\label{eq:m2}
\end{equation}

Using $q=M_2/M_1$ and equation \ref{eq:m2} we obtain the semi-major axis in light-second:

\begin{equation}
a_x =\frac{q}{c}\big[\frac{P_{\rm orb}^2}{4 \pi^2}\frac{G M_{\rm WD}}{(1+q)^2}\big]^{1/3}~{\rm light-second}
\end{equation}
where $c$ is the speed of light. Thus, we finally obtain $\Delta P_{\rm spin}$ as:
\begin{equation}
\Delta P_{\rm spin} =\frac{2 \pi P_{\rm spin}}{P_{\rm orb}}a_x \sin{i}~\rm{s}
\end{equation}

Using the measurements of \citet{mor02gkper} and \citet{sul16GKPer} ($q=0.55\pm0.21$,
$M_{\rm WD}\ge0.86\pm0.02$ M$_{\sun}$, $P_{\rm orb}=1.9968\pm0.0008$ d) we find that 
$a_x \sin{i}=6.1\pm1.8 \sin{i}$ lt-sec and $\Delta P_{\rm spin}=0.078\pm0.023 \sin{i}$ s.
Since the binary inclination lies within the range 50--73$^\circ$ \citep{mor02gkper},
the result is $\Delta P_{\rm spin}=0.042-0.096$ s. 
Thus, the orbital modulation of the spin period significantly affects the measurements of
the latter if the observations last for a shorter time than the orbital period. 
In fact, in fig. \ref{fig:timing} we show that the spin period derived from the 
{\sl Chandra} observation, which is the shortest one, is measured to be longer.

In order to explore a possible spin period variation with the orbital period in our data 
we combined the {\sl NuSTAR} and the {\sl Chandra } HETG 1--6 {\AA} light curves, both 
covering half of the orbital period. Since the light curves show also 
variability on a kilosecond time-scale (see fig. \ref{fig:nuchan_lc})
we fitted and subtracted 5-order polynomial functions from light curve
segments lasting 14 rotation periods each.
The resultant, ``flat'' light curve, was then divided in four parts of equal length 
and the spin period was measured in each part with the PDM method. 

Unfortunately, we only have the ephemerides of \citet{mor02gkper},  
obtained almost 20 years ago, so the error on the phase determination may make it 
non significant. However, the orbital period itself is precisely measured,
and in fig. \ref{fig:oc} we show that $P_{\rm spin}$ is indeed variable. 
Fitting its orbital period dependence with a sine function, we find 
$\Delta P_{\rm spin}=0.050(18)$ s. 
This is close to the lower limit of the expected range of $\Delta P_{\rm spin}$. 
We note, however, that flickering cannot be removed, so the uncertainty in
this measurement is large.

We applied the template fitting method described in \citet{kat09oc} to 
the {\sl NuSTAR} light curves extracted above and below 10 keV in order to measure the
pulse fraction. We folded the light curves with the 351.3255 s period, fitted the mean pulse
profile with a spline function and then used this spline template to fit individual pulses and to 
measure the amplitude and the mean count rate per pulse. The pulse amplitude
was calculated as (fitted pulse maximum - fitted pulse minimum)/2.
The bottom panel of fig. \ref{fig:oc} shows
the result. The correlation between the mean count rate and the amplitude of pulse is very 
prominent and can be fitted with a linear function, providing values of the pulse fraction 
of 65 and 57\% for the soft and hard light curves, respectively. 
Variation of the mean count rate reflects the long term variability on the time-scale of 
kiloseconds (QPOs). The spin modulation is due to the geometric effects and the linear 
dependence between the mean count rate and the pulse amplitude suggests that QPOs in the {\sl NuSTAR}
energy range are due to the intrinsic variability of the emitting source.

\section{Spectral analysis}\label{spec}
\begin{figure*}
\centering
\includegraphics[width=400pt]{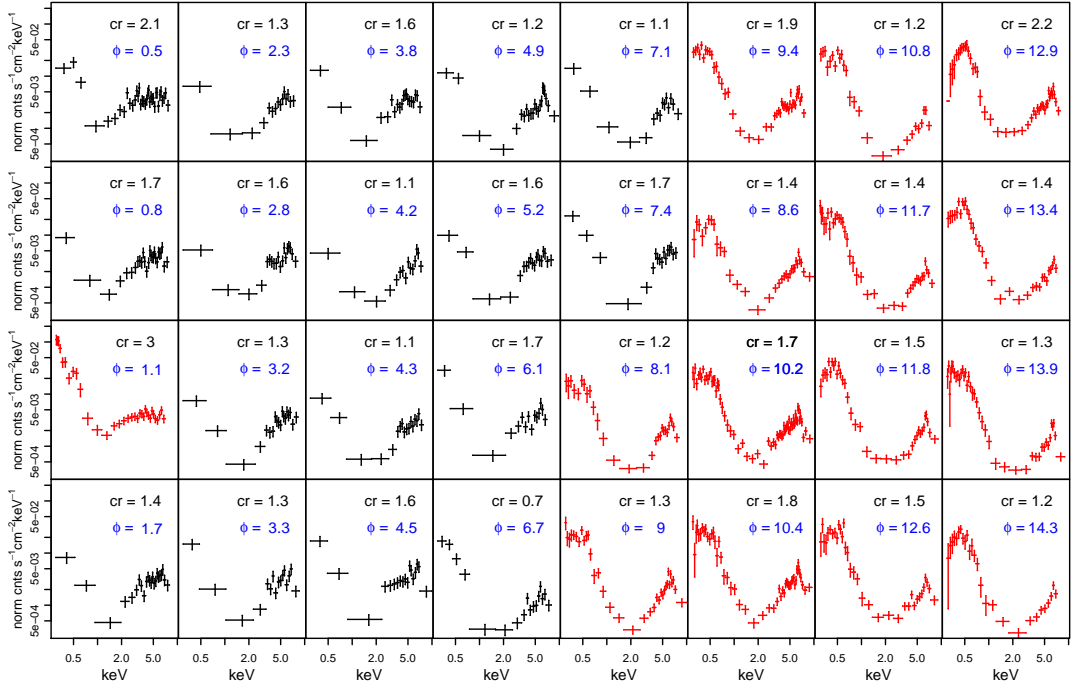}
\caption{Evolution of the Swift XRT spectra of GK Per with time. The black points are data
obtained in the PC mode and the red ones in the WT mode. The mean count rate and the
corresponding orbital phase are marked on each panel.}
\label{fig:swift_spec_tot}
\end{figure*}

The X-ray spectrum of GK Per is very complex. The long-term observations with {\sl Swift} showed
that it is also quite variable, as demonstrated by the hard and soft X-ray light curves in 
fig. \ref{fig:lc}. Fig. \ref{fig:swift_spec_tot} shows the comparison of the {\sl Swift} XRT
spectra obtained on different days, including only spectra with at least 14 data points after 
we binned the PC mode spectra with a minimum of 20 counts per bin, and the WT mode ones with
at least 40 counts per bin. We divided the measurements by the response effective area, so
the WT and PC mode spectra could be directly compared.
Since all the exposures were longer than the spin period, rotation dependent variability 
is smeared out. 
In order to look for possible orbital phase dependence of the spectrum we marked the 
corresponding orbital phase \citep[from the ephemerides in][]{mor02gkper} in each panel.
We also labelled each panel with the mean count rate. We did not find 
significant spectral variability dependent on the orbital phase, although 
all the spectra were different from day to day, sometimes 
with a narrow minimum around 2 keV, other times with a flatter shape around this value. 
Because of these variations there was no possibility to perform a simultaneous fit of the 
{\sl Swift} XRT data and the data from the {\sl Chandra} and {\sl NuSTAR}
observations and we will discuss these observations separately.

From the timing analysis we found that there are at least two different sources of X-ray 
emission in GK Per: one dominates above 2 keV and originates somewhere close to the 
WD, since the flux in this range is modulated with the WD spin period, and the second source,
dominating below 2 keV, is visible during the whole spin cycle. We first analysed the 
hard portion of the X-ray spectrum using the {\sl NuSTAR} and {\sl Chandra} HETG observations.
The {\sl Chandra} HETG spectra, having higher energy resolution provide better 
constrain on the metallicity and the structure of the Fe K complex, while the {\sl NuSTAR}
data allow us to measure the shock temperature.

\begin{figure}
\centering
\includegraphics[width=240pt]{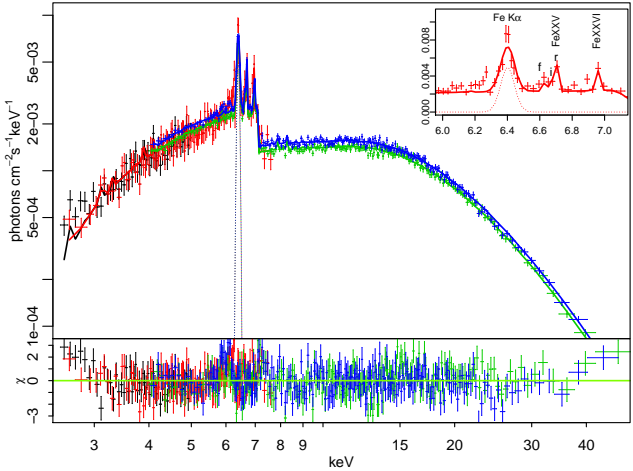}
\caption{The unfolded {\sl NuSTAR} FPMA (blue), FPMB (green), {\sl Chandra} MEG (black)
and HEG (red) spectra and the best-fitting model.
The model components are marked with the dashed lines.
The inset shows the Fe complex in the {\sl Chandra} HEG spectrum.}
\label{fig:nuchan_spec}
\end{figure}

\begin{figure}
\centering
\includegraphics[width=240pt]{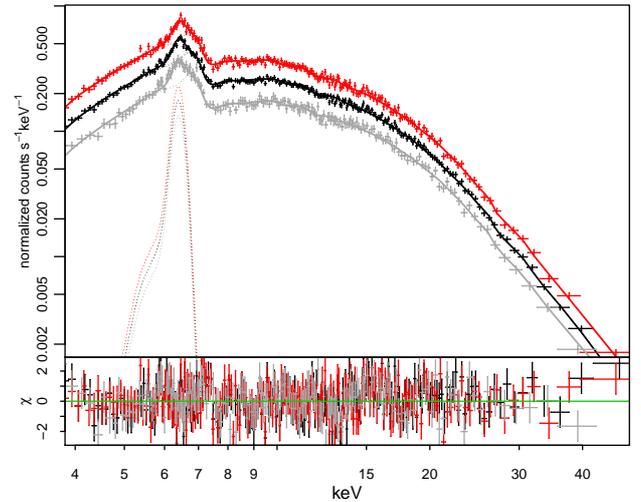}
\caption{The {\sl NuSTAR} FPMA mean (black), on-pulse (red) and off-pulse (grey) 
spectra and the best fitting model.
The model components are marked with the dashed lines.}
\label{fig:nuspec}
\end{figure}

\subsection{The hard X-ray spectral component.}
Hard X-ray emission of accreting magnetic CVs originates in their accretion columns, where
the post-shock plasma is cooling mostly via bremsstrahlung 
radiation and K and L shell line emission as it settles on to the surface of a WD. Therefore, 
for the hard continuum 
we used the cooling flow model \texttt{mkcflow}, which calculates
a plasma in collisional ionization equilibrium (CIE) with a range of temperatures.
\texttt{vmcflow}
model is a modification of the \texttt{mkcflow} with variable abundances of individual elements.
This model was originally created for clusters of galaxies \citep{mus88mkcflow}
but can also be applied to CVs \citep{muk03twotypes}.
The highest plasma temperature (the shock temperature) is an important parameter for 
magnetic CVs, allowing to estimate the WD mass \citep{aiz73WDmass}. 
The emissivity function in the \texttt{mkcflow} model is the inverse of the bolometric luminosity 
 and the model's normalization constant is the mass accretion rate.
In order to investigate the hard X-ray emission, we attempted a simultaneous fit of the 
{\sl NuSTAR} spectra and the {\sl Chandra} HEG+MEG spectra above 2.5 keV.
The {\sl NuSTAR} FPMA and FPMB spectra were fitted together, but the FPMB model was multiplied 
by a constant to account for a slightly different response of the two detectors.
We used the \texttt{vmcflow} model to test the abundances.
The value of the interstellar absorption was obtained from the reddening 
$E(B-V)=0.3$ \citep{wu89GKPerIR} and the nH--$E(B-V)$ relation from \citet{boh78abs}.
The shape of continuum above 2 keV indicates that the emission is highly absorbed, but even with 
partially covering absorber we could not fit the data.
A better result was obtained with the \texttt{pwab} model \citep{don98pwab}, in which the 
fraction of X-rays 
affected by a given column density N(H) is a power-law function of N(H) with index $\beta$. 
We fix the lower temperature of the \texttt{vmcflow} model to the lowest possible value --- 
0.0808 keV because a heavily absorbed spectrum like that of GK Per does not allow
us to determine it accurately, and physically we expect it to be equal to the white
dwarf photospheric temperature.

We also added a Gaussian component to fit the Fe K$\alpha$ fluorescent
line at 6.4 keV. The model's parameters of the best fit are in table \ref{tab:nu_spec}. 
The {\sl NuSTAR} and the hard part of the {\sl Chandra} spectrum together with the described
model are presented in fig. \ref{fig:nuchan_spec}. 
The inset shows the Fe complex measured in the {\sl Chandra} HEG spectrum. The model 
slightly underestimates the flux in the forbidden line of the \ion{Fe}{xxv} triplet, 
which may indicate contribution of the photoionization processes. There are also residuals 
around 6.2--6.3 keV, suggesting Compton-down-scattering of photons. A similar ``shoulder''
of the Fe K$\alpha$ line was detected in previous {\sl Chandra} HETG observations of GK Per
\citep{hel04MCV}.


We used the same procedure of the template fitting, described in the 
previous section, to find the pulse maxima and to extract the on-pulse and off-pulse 
{\sl NuSTAR} spectra. Assuming that the pulse maximum is at $\phi$=0.5, 
we chose time intervals corresponding to the 0.3 -- 0.7 and 0.8 -- 1.2 spin phases and used them 
to extract the on-pulse and off-pulse {\sl NuSTAR} FPMA spectra. 
These spectra are shown in fig. \ref{fig:nuspec} in comparison with the mean one.
All three spectra are remarkably similar. We fitted the 
on-pulse and off-pulse spectra applying the best-fitting model described above, freezing the 
maximum plasma temperature, metallicity and the Fe K$\alpha$ line width.
The best-fitting parameters in table \ref{tab:nu_spec} show that the variation between 
the on-pulse and off-pulse spectra is due to the normalization of the cooling
flow component.

\begin{table*}
\centering
\begin{minipage}{120mm}
\caption{The best-fitting model parameters of the {\sl NuSTAR} mean + {\sl Chandra} HETG, 
on-pulse and off-pulse {\sl NuSTAR} FPMA spectra. 
The model is \texttt{constant$\times$TBabs$\times$pwab$\times$(vmcflow + gaussian)}. 
The errors represent the 90\% confidence region for a single parameter.
We adopted FPMA Constant=1 and FPMB C.=1.089.}
\begin{tabular}{llccc}
\hline
Component & Parameter						&  							\multicolumn{3}{c}{Value}					\\
		  & 								&  	Mean					& On-pulse	& Off-pulse		\\

\hline
TBabs	  & nH ($\times$10$^{22}$ cm$^{-2}$)				& 0.17						& 0.17					& 0.17	\\
\hline
 		  & nH$_{\rm min}$ ($\times$10$^{22}$ cm$^{-2}$) & 7.2$_{-0.4}^{+0.4}$ 		& 7.4$_{-1.3}^{+1.1}$ 	& 7.8$_{-2}^{+2}$\\    
pwab	  & nH$_{\rm max}$ ($\times$10$^{22}$ cm$^{-2}$) & 520$_{-30}^{+30}$ 		& 550$_{-40}^{+40}$ 	& 530$_{-50}^{+50}$ \\    
		  & $\beta$										&-0.199$_{-0.016}^{+0.016}$ & -0.26$_{-0.04}^{+0.05}$& -0.16$_{-0.07}^{+0.07}$	\\
\hline
		  & T$_{\rm low}$ (keV)							& 0.0808					& 0.0808				& 0.0808	\\
		  & T$_{\rm high}$ (keV)						&16.2$_{-0.4}^{+0.5}$		& 16.2					& 16.2		\\
vmcflow   & Fe                  						& 0.105$_{-0.012}^{+0.012}$	& 0.105					& 0.105 		\\  
		  & Ni                  						& 0.1						& 0.1 					& 0.1			\\  
		  & $\dot{m}$$^a$								& 2.6$_{-0.2}^{+0.2}$		& 3.69$_{-0.11}^{+0.12}$& 1.88$_{-0.05}^{+0.09}$\\ 
\hline
		  & E (keV) 									& 6.40      				& 6.40					& 6.40\\
Gaussian  & $\sigma$ (keV)  							& 0.046$_{-0.008}^{+0.012}$	& 0.046					& 0.046	\\
		  & norm ($\times$10$^{-4}$)					& 40$_{-2}^{+2}$ 			& 50$_{-5}^{+5}$		& 33$_{-4}^{+5}$ \\
		  & EW (eV)										& 210$_{-20}^{+30}$			& 191$_{-19}^{+18}$		& 250$_{-200}^{+130}$ \\
\hline 
Flux	  &absorbed$^b$									& 7.24$_{-3.07}^{+0.03}$	& 10.15$_{-0.19}^{+0.05}$& 5.1$_{-0.6}^{+0.6}$\\
Flux	  &unabsorbed$^b$								& 24.2$_{-1.0}^{+1.0}$		& 34.1$_{-1.0}^{+1.1}$	& 17.5$_{-0.5}^{+0.8}$\\
\multicolumn{2}{l}{L ($\times10^{33}$erg s$^{-1}$)} 	& 63$_{-3}^{+3}$			& 90$_{-3}^{+3}$		& 46.1$_{-1.3}^{+2.1}$\\
\multicolumn{2}{l}{$\chi^2$}							& 1.3 						&	\multicolumn{2}{c}{1.0}	\\
     \hline
\multicolumn{5}{p{.9\textwidth}}{{\bf Notes}: $^a$mass accretion rate $\times$10$^{-8}$M$_\odot$ yr$^{-1}$.
$^b$Absorbed and unabsorbed fluxes $\times10^{-10}$ergs cm$^{-2}$ s$^{-1}$ in the 2.5--79 keV energy range. 
The unabsorbed flux was calculated with the \texttt{cflux} command in \texttt{xspec}. }\\
\multicolumn{5}{p{.9\textwidth}}{We assumed a 470 pc distance.}\\
\hline 
\end{tabular}
\label{tab:nu_spec}
\end{minipage}
\end{table*}
  
\subsection{The {\sl Chandra} observation.}\label{ssec:chan}
We analysed the spectrum below 2 keV focusing on the {\sl Chandra} HETG data to investigate
the emission
lines ratio and to derive conclusions about the plasma temperature and density and about the mechanism
of ionisation. Useful indexes of the plasma properties are the {\it R}=$f/i$ and 
{\it G}=$(f+i)/r$ ratios, where $r$, $i$ and $f$ are the fluxes in the resonance, intercombination 
and forbidden lines of the He-like triplets, respectively, and the ratios of H-like to He-like resonance lines
of the same species \citep{gab69triplets}.

\begin{figure}
\centering
\includegraphics[width=240pt]{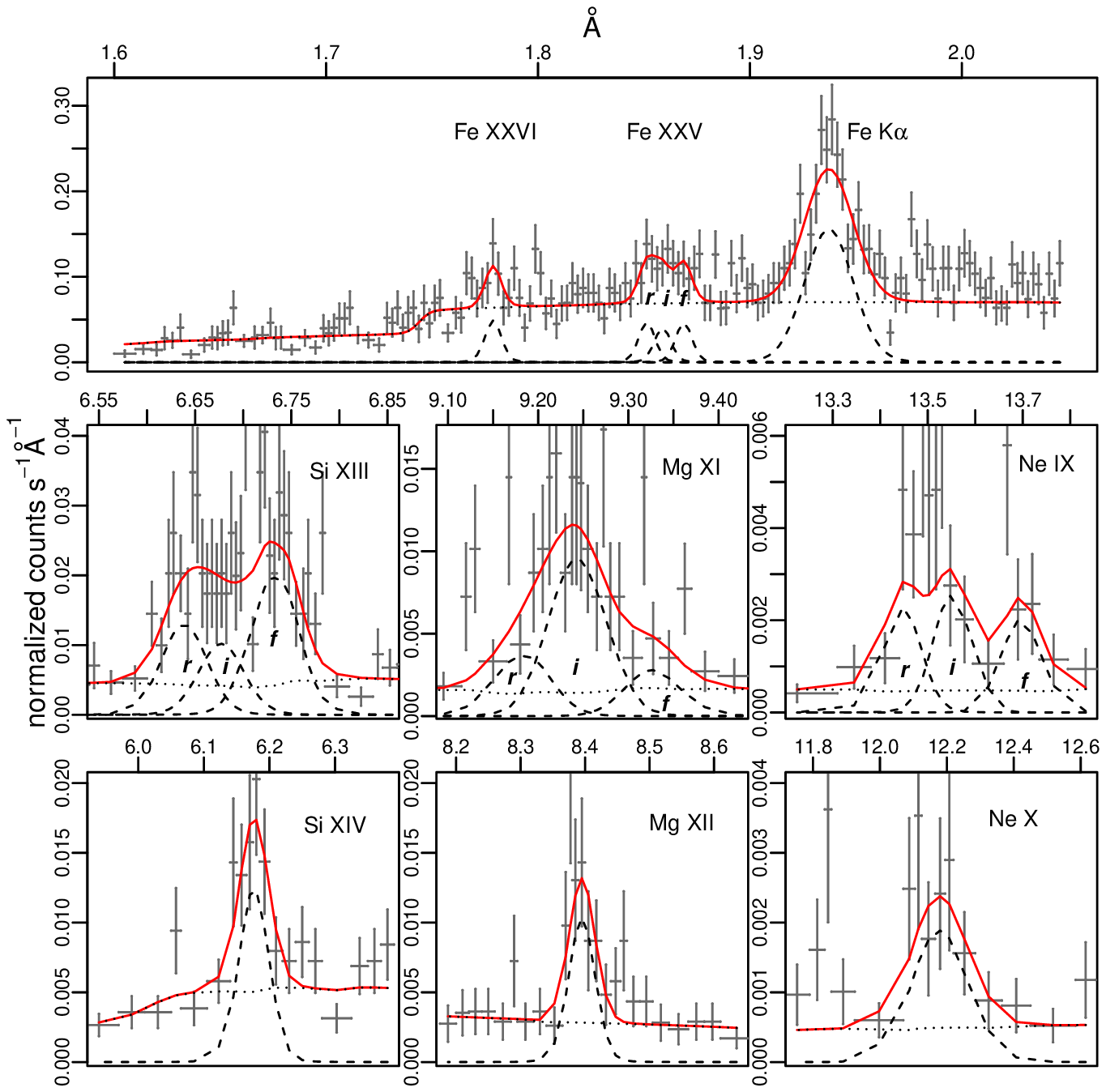}
\caption{Gaussian fits of the strongest emission lines of the {\sl Chandra} HEG (top) and
MEG spectra. In the fits we assumed that the value of $\sigma$ is the same within a 
triplet. The distances between the lines in a triplet were fixed to the table values.
The values of $\sigma$ of the \ion{Fe}{xxvi} and \ion{Fe}{xxv}$_{\rm r,f,i}$ lines were fixed 
to the instrument resolution.}
\label{fig:lines}
\end{figure}

We fitted all the emission lines in the {\sl Chandra} MEG spectrum with Gaussians, 
assuming that the Ne, Mg and Si lines are absorbed only by the interstellar absorption, 
and we used a power law to represent the level of continuum. We also assumed that 
the widths of the lines within a triplet were constant and fixed the distances between 
these lines to the table values.
For the Gaussian fit of the Fe K complex in the {\sl Chandra} MEG we also introduced the \texttt{pwab} model 
with the parameters of the fit of the mean {\it NuSTAR} spectrum and a bremsstrahlung 
component at kT=14 keV for the underlying continuum. 
We did not find any significant line shifts departing from the laboratory wavelengths.
The resulting broadening
and flux are presented in table \ref{tab:lines} and in fig. \ref{fig:lines}. 
We also measured the {\it R} and {\it G} ratios for the Si, Mg and 
Ne triplets, and give them in table \ref{tab:rg}. 

\begin{figure*}
\centering
\includegraphics[width=450pt]{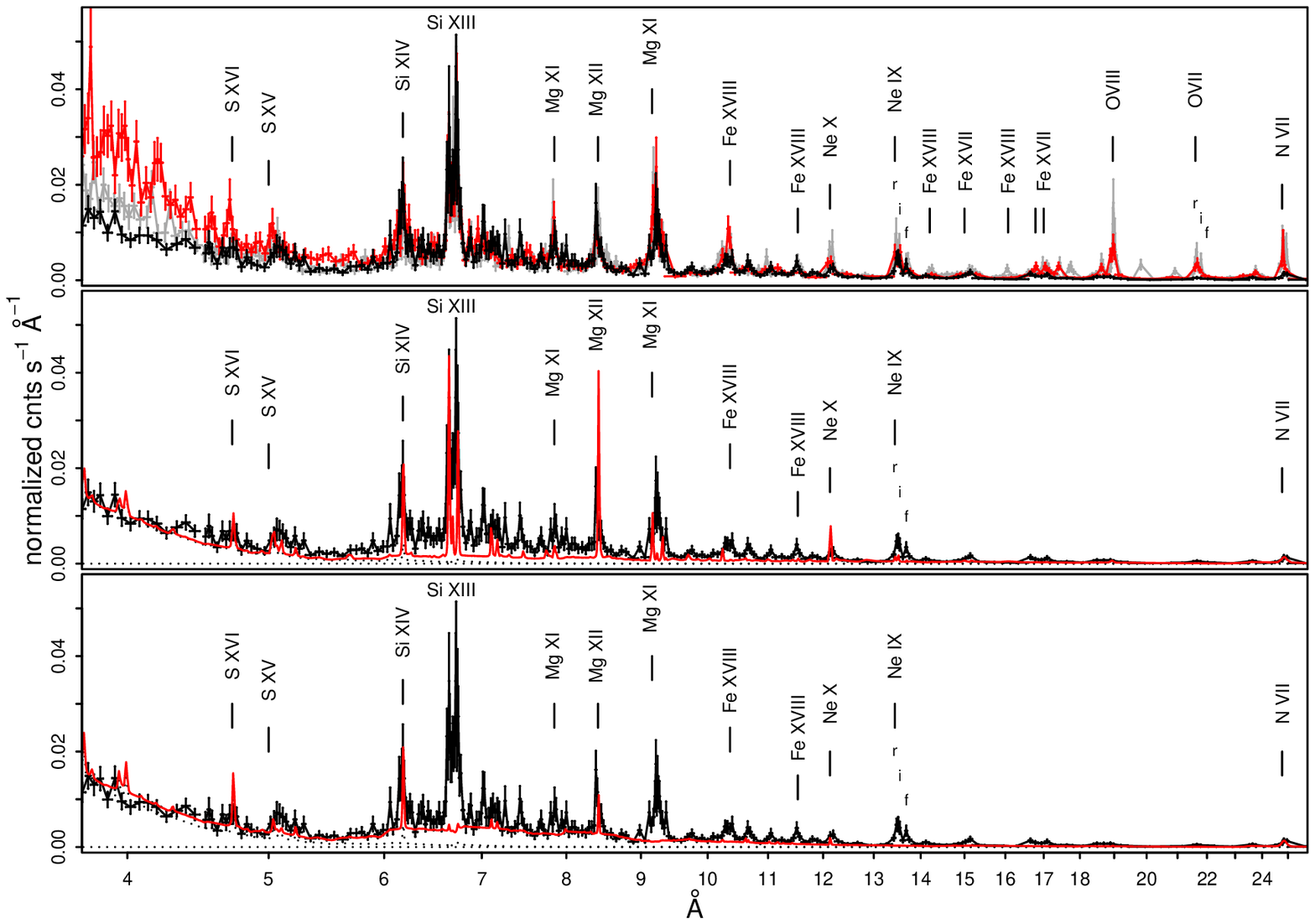}
\caption{From top to bottom: comparison of the {\sl Chandra} MEG spectra of GK Per in outbursts in 2015 and 2002.  
The {\sl Chandra} MEG spectrum discussed in this paper is plotted in black, while the Chandra MEG 
spectra obtained on March 27 and April 9 2002 (PI: C. Mauche) are plotted in red and grey, 
respectively. All the observations were performed close to the optical maxima. Middle panel: the 2015 
{\sl Chandra} MEG spectrum and the \texttt{TBabs$\times$(pwab$\times$(vmcflow + gaussian) + vapec + bb + gaussian)}
model (the red line). The temperature of the \texttt{vapec} component was fixed to 0.9 keV. Bottom panel:
The 2015 {\sl Chandra} MEG spectrum with the same model and the 4.9 keV temperature of the \texttt{vapec}.}
\label{fig:comp}
\end{figure*}

The {\it G} ratio is around 2, which means that there is no strong photoionizing component (in the case of
pure photoionized plasma G is $\sim$4). We either have a collisional-ionization mechanism
or a `hybrid plasma', a mixture of collisional and photoionization \citep{por00photoion}.
The He-like triplets of different elements show very different line ratios. The \ion{Si}{xiii} 
triplet has a very strong forbidden line, which cannot be explained solely with collisional 
ionization, while \ion{Mg}{xi} and \ion{Ne}{ix} have quite weak forbidden lines, but very 
strong intercombination lines, which indicates a high density \citep{por00photoion}.

\begin{table*}
\centering
\begin{minipage}{170mm}
  \caption{Emission lines broadening, shifts and fluxes in the {\sl Chandra} HETG spectra.}
  \begin{tabular}{lllllccc}
  \hline
Line			 &E$_{\rm rest}$& E$_{\rm max}$ 	  &$\Delta\upsilon$ 	& $\sigma$			  &F$_{\rm abs}$$\times10^{-13}$ & F$_{\rm unabs}$ $\times10^{-5}$ & F$_{\rm unabs}$$\times 10^{-13}$ \\
				 &keV			& keV 		 		  &	km s$^{-1}$			&km s$^{-1}$		  &ergs cm$^{-2}$ s$^{-1}$		 & ph cm$^{-2}$ s$^{-1}$ 		   & ergs cm$^{-2}$ s$^{-1}$\\
\hline
Ne X			 & 1.02195 & 1.017$^{+0.004}_{-0.004}$&-1400$^{+1200}_{-1200}$&2100$^{+1000}_{-800}$	&0.6$^{+0.2}_{-0.2}$	& 5$^{+2}_{-2}$			& 0.9$^{+0.3}_{-0.4}$\\
Ne IX$_{\rm r}$  & 0.92200 & 						  &						& 					  		&0.5$^{+0.3}_{-0.3}$	& 6$^{+3}_{-3}$			& 0.8$^{+0.5}_{-0.5}$\\
Ne IX$_{\rm i}$  & 0.91488 & 		 				  &						& 800$^{+600}_{-300}$ 		&0.6$^{+0.3}_{-0.3}$	& 7$^{+4}_{-4}$    		& 1.0$^{+0.5}_{-0.5}$\\
Ne IX$_{\rm f}$  & 0.90510 & 						  &						&  					  		&0.5$^{+0.3}_{-0.3}$	& 6$^{+4}_{-4}$    		& 0.9$^{+0.5}_{-0.5}$\\
\hline 
Mg XII			 & 1.47264 & 1.477$^{+0.003}_{-0.002}$& 900$^{+600}_{-400}$	&700$^{+500}_{-300}$		&0.2$^{+0.1}_{-0.1}$	& 1.3$^{+0.4}_{-0.4}$	& 0.3$^{+0.1}_{-0.1}$\\
Mg XI$_{\rm r}$  & 1.35225 & 						  &						& 					  		&0.3$^{+0.2}_{-0.2}$	& 0.8$^{+0.9}_{-0.8}$	& 0.4$^{+0.2}_{-0.2}$\\
Mg XI$_{\rm i}$  & 1.34332 & 						  &						&1100$^{+1800}_{-800}$ 		&0.7$^{+0.3}_{-0.2}$	& 4.9$^{+2.3}_{-1.4}$	& 0.9$^{+0.3}_{-0.3}$\\
Mg XI$_{\rm f}$  & 1.33121 & 						  &						& 					  		&0.16$^{+0.12}_{-0.15}$	& 1.1$^{+0.7}_{-0.6}$	& 0.20$^{+0.15}_{-0.19}$\\
\hline 
Si XIV			 & 2.00608 & 2.008$^{+0.002}_{-0.002}$& 300$^{+300}_{-300}$	&800$^{+400}_{-300}$ 	  	&0.51$^{+0.16}_{-0.15}$ & 1.7$^{+0.5}_{-0.5}$	& 0.56$^{+0.17}_{-0.17}$\\
Si XIII$_{\rm r}$& 1.86500 & 1.8674$^{+0.0029}_{-0.0015}$& 400$^{+500}_{-200}$	& 					  	&0.62$^{+0.11}_{-0.25}$ & 1.9$^{+0.8}_{-0.5}$	& 0.7$^{+0.2}_{-0.2}$\\
Si XIII$_{\rm i}$& 1.85423 & 1.8566$^{+0.0029}_{-0.0015}$& 400$^{+500}_{-200}$	&1000$^{+1200}_{-700}$ 	&0.34$^{+0.2}_{-0.19}$	& 1.5$^{+0.7}_{-0.8}$	& 0.4$^{+0.2}_{-0.2}$\\
Si XIII$_{\rm f}$& 1.83967 & 1.8421$^{+0.0029}_{-0.0015}$& 400$^{+500}_{-200}$	& 					  	&0.77$^{+0.16}_{-0.2}$	& 2.9$^{+0.6}_{-0.8}$	& 0.8$^{+0.2}_{-0.2}$\\
\hline 
Fe XXVI			 & 6.97316 & 	 				      &						& 					  		&8$^{+4}_{-4}$  	&  50$^{+20}_{-20}$& 50$^{+20}_{-20}$\\
Fe XXV$_{\rm r}$ & 6.70040 & 	 				      &						& 					  		&6$^{+4}_{-4}$	  	&  40$^{+20}_{-20}$& 40$^{+30}_{-30}$\\
Fe XXV$_{\rm i}$ & 6.67000 & 	 				      &						& 					  		&5$^{+4}_{-4}$  	&  40$^{+20}_{-20}$& 40$^{+30}_{-30}$\\
Fe XXV$_{\rm f}$ & 6.63659 & 	 				      &						& 					  		&6$^{+3}_{-3}$	  	&  40$^{+20}_{-20}$& 40$^{+20}_{-20}$\\
Fe K$_{\alpha}$  & 6.40384 & 6.400$^{+0.007}_{-0.007}$& -200$^{+300}_{-300}$&1600$^{+300}_{-300}$ 		&48$^{+6}_{-6}$	  	& 350$^{+50}_{-50}$& 360$^{+50}_{-50}$ \\
\hline 
\multicolumn{8}{p{.9\textwidth}}{{\bf Notes:} We assumed that the value of $\sigma$ is the 
same within a triplet for the Ne, Mg and Si lines.
The distances between the lines in a triplet were fixed to the table values.
The values of $\sigma$ of the \ion{Fe}{xxvi} and Fe {Fe}{xxv}$_{\rm r,f,i}$ lines were frozen 
to the instrumental spectral resolution. The absorbed and the unabsorbed fluxes in ergs cm$^{-2}$ s$^{-1}$ were calculated using the \texttt{cflux} command. The 
unabsorbed flux in ph cm$^{-2}$ s$^{-1}$  was calculated from the normalization constant of the Gaussian model.}\\
\hline 
\end{tabular}
\label{tab:lines}
\end{minipage}
\end{table*}

\begin{table}
\centering
\begin{minipage}{80mm}
  \caption{{\it R}, {\it G} and H-like/He-like$_{\rm r}$ ratios.}
  \begin{tabular}{lccc}
  \hline
Element	&{\it R}$=f/i$  &{\it G}$=(f+i)/r$&  H-like/He-like$_{\rm r}$\\
\hline
Ne 		& 0.9$\pm$0.7 	& 2.4$\pm$1.7 & 1.1$\pm$0.8\\
Mg 		& 0.2$\pm$0.2 	& 2.8$\pm$1.6 &	0.8$\pm$0.5\\
Si 		& 2.0$\pm$1.1 	& 1.7$\pm$0.6 &	0.8$\pm$0.3\\
Fe 		& 1.0$\pm$0.7 	& 2.0$\pm$1.2 &	1.3$\pm$0.8\\
\hline 
\multicolumn{4}{p{.9\textwidth}}{{\bf Notes}: the values of {\it R} and {\it G} were 
calculated from the values of the unabsorbed flux from table \ref{tab:lines}.}\\
\hline 
\end{tabular}
\label{tab:rg}
\end{minipage}
\end{table}

\citet{muk03twotypes} showed that the {\sl Chandra} HETG spectrum of GK Per obtained during the outburst
in 2002 is consistent with the predictions of a photoionization model with a power law as 
the photoionizing continuum. Although the power law emission gave a good approximation, and
could indeed photoionize the plasma in which the soft X-rays emission lines are produced, 
such a non thermal component in a CV does not seem to have a physical reason. Given also the G ratio
and the absence of a clearly non-thermal component in the {\sl NuSTAR} range, we
do not favour this explanation for the present set of observations.

The top panel of figure \ref{fig:comp} shows the comparison of the {\sl Chandra} MEG spectrum obtained in 2015 
with {\sl Chandra} MEG data discussed in \citet{muk03twotypes}. The most
recent spectrum of GK Per has much weaker lines in the region above 20 {\AA}, which is due to 
contaminant build-up of the {\sl Chandra} HETG+ACIS detector. The low energy effective area
is reduced in 2015 compared to 2002. The
intensities of the lines that are in the 6--11 {\AA} region are almost the same.
The {\sl Chandra} spectra in fig. \ref{fig:comp} give an additional proof that there
are several distinct sources of emission: there is no correlation between the 6--11{\AA} emission
lines strengths and the hard continua below 5 {\AA}.

\begin{figure}
\centering
\includegraphics[width=220pt]{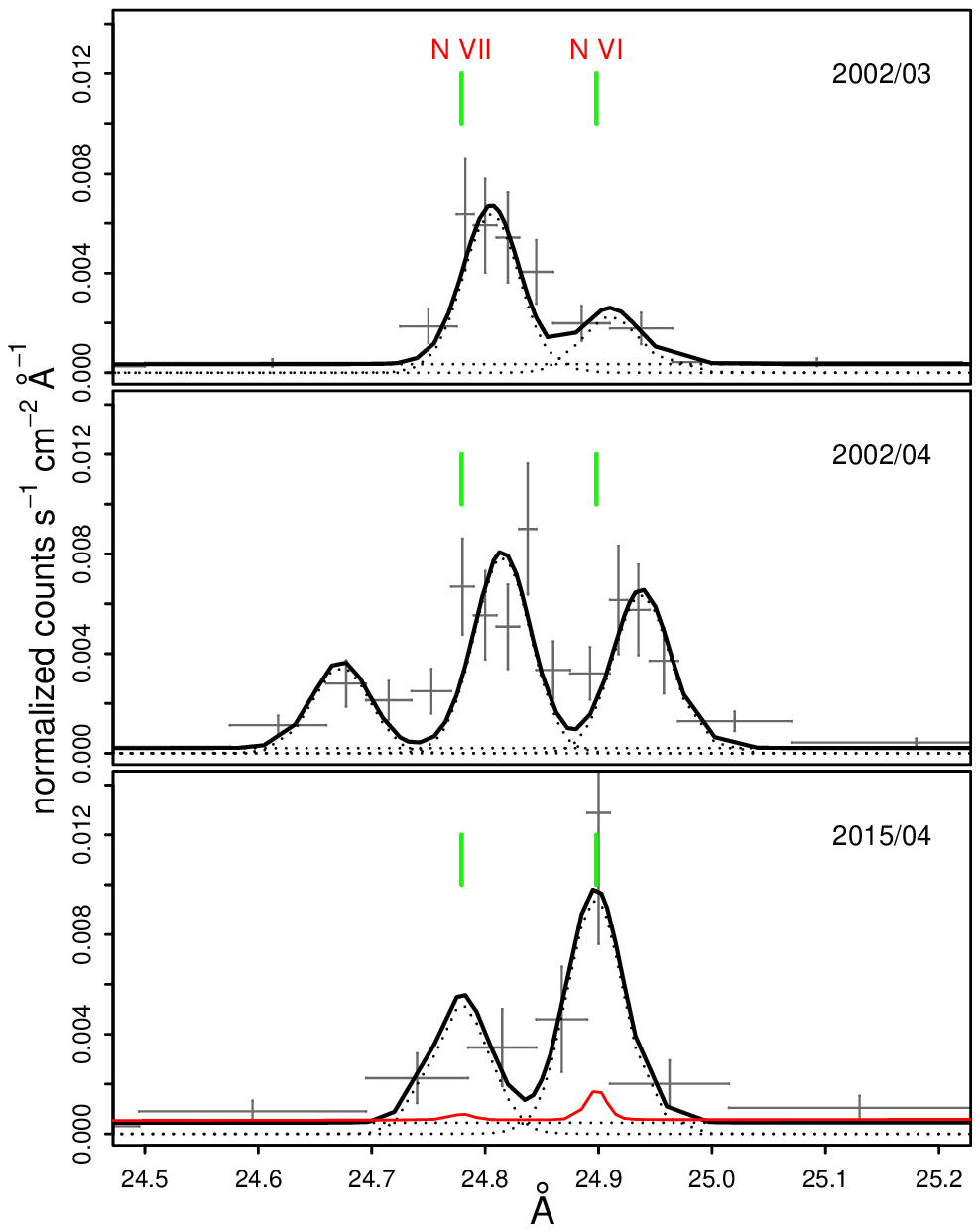}
\caption{Comparison of the \ion{N}{vii} emission line's profile measured with the {\sl Chandra} MEG 
in 2015 and 2002 (the data sets are the same as in fig. \ref{fig:comp}). The dates of observations 
and the rest-frame positions of the \ion{N}{vii}
and \ion{N}{vi} He$\beta$ lines are marked at each panel. All the lines were fitted with two or three 
Gaussians and a power law to represent the continuum level. In the bottom panel we also 
show the contribution from the nova shell, based on the model of \citet{tak15GKPershell} and
a power law to fit the underlying continuum.}
\label{fig:nvii}
\end{figure}

The \ion{N}{vii} line detected in the {\sl Chandra} MEG 2002 and 2015 spectra showed 
remarkably different profiles in comparison to the other lines. The \ion{N}{vii} line's regions
are shown in fig. \ref{fig:nvii}. In March 2002 and April 2015 two different 
emission lines were resolved around the rest-frame position of \ion{N}{vii},
while in April 2002 another blue-shifted component could be distinguished. Table \ref{tab:nvii} 
shows the central positions and the unabsorbed fluxes of all the lines that were resolved
in three {\sl Chandra} MEG spectra. \citet{vie05GKPer} in their Reflection Grating Spectrograph (RGS) spectrum of GK Per 
also noticed that the \ion{N}{vii} line had a different structure that could be approximated by three 
Lorentzian profiles. Two scenarios could explain these emission lines. It can be one central
\ion{N}{vii} line and one/two \ion{N}{vii} lines shifted from the main line. Another possibility
is that the line around 24.8 {\AA} is indeed the \ion{N}{vii} emission,
but the line around 24.9 {\AA} is the \ion{N}{vi} He$\beta$ line, both with zero velocity.
From the {\sl Chandra} MEG spectra we find that if the lines 
around 24.8 {\AA} are different components of the central \ion{N}{vii} line, this would indicate
velocity shifts of 1200 -- 1600 km s$^{-1}$. Similar red and blue-shifted emission line's components,
although with smaller velocity shifts, were measured in the optical spectra of GK Per during
outbursts \citep{bia03GKPershift} and were attributed to the emission from the matter 
in the magnetosphere, falling on to the WD. A question to answer is why \ion{N}{vii}
is the only line that shows such a complex profile. Two {\sl Chandra} MEG spectra
obtained near optical maximum of the same outburst (see tab. \ref{tab:nvii}) also clearly show that
the flux of the different \ion{N}{vii} line's components varies with time, while the flux of 
central line is stable within the errors. 
If we observe, instead, the \ion{N}{vii} and \ion{N}{vi} He$\beta$ emission lines the 
relative intensities of the two lines can be used as temperature indicator 
(under the assumption of collisional ionization equilibrium). The different relative intensities
in March and April of 2002 would imply that the temperature changes. The very strong 
\ion{N}{vi} He$\beta$ line in comparison to \ion{N}{vii} in the 2015 spectrum indicates 
also a very low plasma temperature: $\lesssim$0.08 keV.

We checked whether the expanding nova shell can contribute to line emission in soft X-rays. 
\citet{bal05GKPerShell} showed that the shell has enhancement in the elemental abundances 
of Ne and N and \citet{vie05GKPer} fitted the quiescence Chandra ACIS-S spectrum of the shell
with pure emission lines of N, O and Ne, although the lines could not be resolved.
In order to estimate a possible contamination of the \ion{N}{vii} line from the central source
by the shell emission, we compared the predictions of the shell emission model from \citet{tak15GKPershell} 
with the 2015 {\sl Chandra} MEG spectrum. The model from \citet{tak15GKPershell} with the 
only modification in the N abundance (we assumed N/N$_{\sun}$=5) and a power law to represent the continuum
level is shown with the red line in the bottom panel of fig. \ref{fig:nvii}.
We see that the \ion{N}{vii}$/$\ion{N}{vi} line's flux from the entire shell is much smaller
than that measured in the 2015 {\sl Chandra} MEG spectrum.
We conclude that the N lines around 24.8 {\AA} originate not from the extended shell, but 
the nature of the different components is unknown.

\begin{table*}
\centering
\begin{minipage}{170mm}
  \caption{\ion{N}{vii} emission line fluxes, measured in the {\sl Chandra} MEG spectra of 2002 and 2015.}
  \begin{tabular}{lccccccccc}
  \hline
			  & \multicolumn{3}{c}{central}		& \multicolumn{3}{c}{red-shifted} 		& \multicolumn{3}{c}{blue-shifted} \\
Observation   & E max (keV) 				& Flux$^*$ & Ph. flux$^{**}$ &  E max & Flux & Ph. flux &  E max & Flux & Ph. flux\\
\hline
2002 March 27 & 0.4998$^{+0.0003}_{-0.0003}$& 12$\pm$4 & 1.6$\pm$0.4 & 0.4977$^{+0.0006}_{-0.0008}$ & 5$\pm$2 	& 0.6$\pm$0.3 & & & \\
2002 April 9  & 0.5000$^{+0.0003}_{-0.0004}$& 11$\pm$3 & 1.4$\pm$0.4 & 0.4973$^{+0.0005}_{-0.0003}$ & 11$\pm$3	& 1.3$\pm$0.4 & 0.5026$^{+0.0002}_{-0.0005}$ & 5$\pm$2& 0.6$\pm$0.3 \\
2015 April 4  & 0.50036 		 			& 9$\pm$6  & 1.1$\pm$0.7 & 0.4980 						& 17$\pm$8 	& 1.2$\pm$1.0 & & & \\
\hline 
\multicolumn{10}{p{.9\textwidth}}{{\bf Notes}: $^*$Unabsorbed flux$\times 10^{-13}$ergs cm$^{-2}$ s$^{-1}$. 
$^{**}$Unabsorbed photon flux $\times 10^{-3}$ photons cm$^{-2}$ s$^{-1}$.
The fluxes were measured from the Gaussian fits, 
shown in fig. \ref{fig:nvii} using \texttt{cflux} and \texttt{cpflux} commands. All the line widths were
fixed to the value of the instrumental resolution. In these fits 
we assumed the same value of the interstellar absorption as in the rest of the paper: 0.17$\times$10$^{22}$cm$^{-2}$.}\\
\hline 
\end{tabular}
\label{tab:nvii}
\end{minipage}
\end{table*}

The complexity of the spectrum is demonstrated by the ratios of H to He-like lines, which in the 
case of pure collisional ionization is a signature of the plasma temperature. 
The He-like lines are stronger than the H-like for all the species. The \ion{Si}{xiii}
to \ion{Si}{xiv} lines ratio indicates a temperature $\sim$0.9 keV, so  
the origin of these lines is not in the hotter plasma that explains the {\sl NuSTAR} spectrum.
The He to H-like lines ratios of Mg and Ne correspond to even lower plasma
temperatures.
These lines also cannot be explained by another \texttt{mkcflow} component at lower temperature
since the cooling flow model always produces H-like lines stronger than the He-like lines \citep{lun15EXHyd}. 
The middle and bottom panels of fig. \ref{fig:comp} show the comparison of the {\sl Chandra}
MEG spectrum with the predictions of the single temperature thermal plasma emission model. 
We added to the best-fitting model of the {\sl NuSTAR} data a \texttt{vapec} component 
(a single-temperature plasma in CIE with variable abundances of individual elements) and
a Gaussian at 0.5 keV to represent the \ion{N}{vii} line. Following \citet{vie05GKPer} and
 \citet{eva09GKPer} we also introduced a blackbody
component to represent the thermalized X-ray emission from the WD surface at kT = 66 eV. 
The choice of the blackbody temperature will be explained in the next section. 
In the middle panel of fig. \ref{fig:comp} the 
temperature of the \texttt{vapec} component was fixed to 0.9 keV, in order to fit the \ion{Si}{xiii}
to \ion{Si}{xiv} lines ratio. In this case the model underestimates the level of
continuum and overestimates the He to H-like lines ratios of Mg and Ne. In the bottom panel
the temperature of the \texttt{vapec} component corresponds to the best-fitting value --- 4.9 keV, 
which correctly estimates the level of continuum, but cannot reproduce the line ratios.
A lower temperature \texttt{vapec} component with higher normalization constant affected 
by a complex absorber could explain the Si lines
and the continuum level, but not the He to H-like lines ratio of Mg and Ne. The emission
lines ratios clearly indicate a multi-temperature plasma emission. However, we added another \texttt{apec} 
component to fit the lines at longer wavelengths, but it did not improve the fit significantly.
Fig. \ref{fig:comp} also shows that the \texttt{vapec} model underestimates the 
intercombination and forbidden lines in all the triplets. 
Thus, the overall spectrum below 2 keV cannot be represented with a model of plasma
in CIE, not with a cooling flow, neither with single or two-temperature \texttt{vapec} model.

Fig. 8 in \citet{por00photoion} shows that the temperature of 0.9 keV and the value 
of the R ratio of the Si triplet (see table \ref{tab:rg}), which is a density indicator,
corresponds to the electron density n$_{\rm e}\sim$3$\times$10$^{13}$ cm$^{-3}$.
Using these estimates and the value of normalization of the \texttt{vapec} component (which gives
the emission measure) we find that the radius of the emitting source is 
$\sim 1.4$R$_{\rm WD}$, assuming a spherical distribution of the emitting plasma, 
M$_{\rm WD}$=0.86 M$_{\sun}$ and a WD mass-radius relation from \citet{nau72WDrad}. 
It should be mentioned, however, that the G and R ratios are measured with large uncertainty
(see table \ref{tab:rg}) and we cannot evaluate the contribution of photoionization 
processes, so this is a qualitative estimate.

\begin{table*}
\centering
\begin{minipage}{150mm}
\caption{Model parameters used for the fit of {\sl Chandra} MEG spectrum apart from that listed
in the first column of table \ref{tab:nu_spec}. 
The model is \texttt{TBabs$\times$(pwab$\times$(vmcflow + gaussian) + bb + gaussian)}.
The fit was performed for two different temperatures.
The parameters without errors were fixed to the values in this table.
The Fe and Ni abundances were fixed to 0.105 and 0.1, respectively.}
\begin{tabular}{llcc}
\hline
Component & Parameter							&  \multicolumn{2}{c}{Value} \\
\hline
vapec  	 & T (keV) 								& 0.9							&	4.9$_{-0.8}^{+1.4}$	\\
		 & norm ($\times$10$^{-3}$) 			& 1.1$_{-0.1}^{+0.1}$			&	2.6$_{-1.5}^{+1.5}$\\
\hline 
bb		  & T (eV) 								& 66 							&	66 	\\
		  & norm ($\times$10$^{-4}$)			& 13$_{-2}^{+2}$				&	12$_{-2}^{+2}$	\\
\hline
		  & E (keV) 							& 0.4988$_{-0.0009}^{+0.0011}$	&	0.4988$_{-0.0009}^{+0.0011}$  \\
Gaussian  & $\sigma$ (keV)  					& 0.0019$_{-0.0009}^{+0.0011}$	&	0.0019$_{-0.0009}^{+0.0011}$ \\
		  & norm ($\times$10$^{-4}$)			& 28$_{-12}^{+13}$				&	29$_{-12}^{+13}$	\\

\hline 
\end{tabular}
\label{tab:global}
\end{minipage}
\end{table*}

\subsection{{\sl Swift} XRT observations}

\begin{table}
\centering
\begin{minipage}{80mm}
\caption{The best fitting model parameters of the {\sl Swift} XRT data. 
\texttt{TBabs$\times$(pwab$\times$(vmcflow + gaussian) + vapec + bb)}. 
The Fe and Ni abundances of the \texttt{vmcflow} and \texttt{vapec} components 
were fixed to 0.106 and 0.1, respectively.
The errors represent the 90\% confidence region for a single parameter.}
\begin{tabular}{llcc}
\hline
Component & Parameter						&  \multicolumn{2}{c}{Value} 					\\
		  & 								&  first		  & second			\\
		  & 								&  two weeks  	  & two weeks			\\
\hline
TBabs	  & nH$^{a,b}$ 						& 0.17				&	0.17 \\
\hline
 		  & nH$_{\rm min}$$^b$  			& 2.7$_{-1.2}^{+0.6}$ 	& 5.0$_{-1.0}^{+1.1}$ \\    
pwab	  & nH$_{\rm max}$$^b$				& 75$_{-14}^{+18}$ 		& 90$_{-20}^{+20}$ 	\\    
		  & $\beta$$^a$						& 0						& 0			\\
\hline
vmcflow   & T$_{\rm high}$ (keV)			& 17$_{-4}^{+10}$		& 17$_{-4}^{+15}$ \\
		  & $\dot{m}$$^c$					& 0.7$_{-0.3}^{+0.5}$ 	& 0.6$_{-0.3}^{+0.3}$\\ 
\hline
		  & E (keV) 						& 6.4     				& 6.4\\
Gaussian  & $\sigma$ (keV)  				& 0.04					& 0.04\\
		  & norm ($\times$10$^{-4}$) 		& 6$_{-2}^{+2}$			& 15$_{-2}^{+2}$\\
\hline
vapec	  & T (keV) 						&  $>$1.9	     		& 0.80$_{-0.10}^{+0.20}$\\
		  & norm ($\times$10$^{-3}$) 		&  2.5$_{-0.5}^{+3.1}$	& 3.7$_{-0.3}^{+0.3}$\\
\hline 
bb		  & T (eV) 							& 75$_{-3}^{+3}$  		& 63$_{-2}^{+3}$\\
		  & norm ($\times$10$^{-4}$)		& 6.2$_{-0.8}^{+1.0}$	& 36$_{-5}^{+5}$\\
\hline
Flux$_{\rm 0.3-2 keV}$$^d$& abs.    		& 4.6$_{-1.0}^{+0.3}$	& 11.9$_{-0.5}^{+0.2}$ \\
						  & unabs. 			& 420 					& 430\\
Flux$_{\rm 2-10 keV}$$^d$ & abs.    		& 165$_{-60}^{+2}$		& 117$_{-50}^{+2}$\\
						  & unabs. 			& 507					& 451 \\
\multicolumn{2}{l}{L$_{\rm bb}$ ($\times10^{33}$erg s$^{-1}$)} 		& 1.33$_{-0.17}^{+0.20}$& 8.6$_{-0.7}^{+0.7}$	\\
\multicolumn{2}{l}{R$_{\rm bb}$$^g$ ($\times10^6$cm)}				& 1.70$_{-0.12}^{+0.13}$& 5.3$_{-0.3}^{+0.5}$\\
\multicolumn{2}{l}{L$_{\rm 2-10 keV}$ ($\times10^{33}$erg s$^{-1}$)}& 13.4					& 11.9					\\
\multicolumn{2}{l}{$\chi^2$}				& \multicolumn{2}{c}{1.7}					\\
\hline
\multicolumn{4}{p{.9\textwidth}}{{\bf Notes}: $^a$Frozen parameter. $^b$$\times$10$^{22}$ cm$^{-2}$.}\\
\multicolumn{4}{p{.9\textwidth}}{$^c$Mass accretion rate $\times$10$^{-8}$M$_\odot$ yr$^{-1}$.
$^d$$\times10^{-12}$erg cm$^{-2}$ s$^{-1}$.}\\
\multicolumn{4}{p{.9\textwidth}}{$^g$ Radius of the emitting region. We assumed a 470 pc distance.}\\
\hline 
\end{tabular}
\label{tab:swift}
\end{minipage}
\end{table}

\begin{figure}
\centering
\includegraphics[width=220pt]{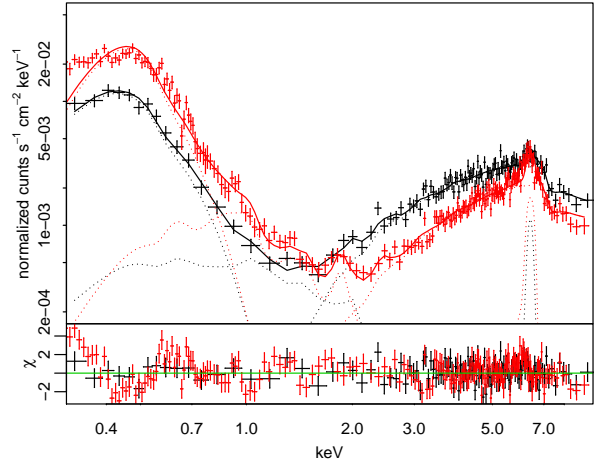}
\caption{The averaged {\sl Swift XRT} spectra obtained during the first (black) and the 
second two weeks (red) of the observations and the best-fitting model. The model
parameters are  plotted with the dotted lines.}
\label{fig:weeks}
\end{figure}

Fig. \ref{fig:weeks} shows the comparison of the average {\sl Swift} XRT spectra in the first
two weeks and in the following two weeks. The soft flux increased with time, while the hard
flux decreased. We fitted the average spectra with the model described in section \ref{ssec:chan}
in order to estimate the changes of the flux in different spectral regions.
There was no possibility to measure the Fe K$\alpha$ line width in the {\sl Swift} spectra,
so we fixed its central position and the $\sigma$ to the values from tab. \ref{tab:nu_spec}.
We assumed that the metallicity did not change with time and the corresponding parameters 
in the \texttt{vapec} component were constrained to be the same in the two data sets.
The best-fitting parameters are listed in the table \ref{tab:swift}. Here and in the 
following sections we refer to
the spectral regions as soft (blackbody-like, below 0.8 keV),
intermediate (between 0.8 and 2 keV) and hard (cooling flow component, above 2 keV).
There are significant residuals in the part of the spectra below 2 keV and in particular
below 0.4 keV in the spectrum obtained in the later period. There is also 
an excess around 0.9 keV, where the \ion{Ne}{ix} line is.
 
Although the model is quite approximate, we concluded that the hard X-ray flux decreased
mostly because of the increased absorption. The soft X-ray flux increased, first of all
because of the increased normalization of the blackbody component and the
\texttt{vapec} components. 
The blackbody emitting area increased by a factor of 3, but it remained of 
of the order of $10^{-5}-10^{-6}$ the area of the WD surface, which is the typical
size of a heated polar region in soft IPs \citep[see e.g.][]{ber12newIP}.

\section{Discussion}

\subsection{The WD spin and the long-term variations.}
The {\sl NuSTAR} observations of GK Per provided the first detection of a high amplitude 
modulation due to the WD spin period in X-rays above 10 keV in an IP (only XY Ari in 
outbursts is known to show a comparable amplitude of modulation).
The fact that the spin modulation is so strong in hard X-rays and that the pulse amplitude 
is not energy dependent indicates that the modulation is a geometric effect rather than
due to absorption as in the majority of IPs. This modulation can be partially explained 
by an obscuration of the lower accretion pole by the inner disc \citep{hel97XYAri, hel04GKPerRXTE}. However,
the obscuration of the lower pole alone, does not explain the pulse profile. 
A small shock region with a low shock height will either be completely visible or completely 
behind the white dwarf with very little transition in between, resulting in a 
square wave spin modulation. In the case of GK Per the modulation is not a square
wave but quasi-sinusoidal and about 40$\%$ of the total flux is always visible, suggesting
a large shock height or an extended shock region. 
In the first case the soft X-rays, originating closer to the WD surface, will show more 
prominent modulation, while the hardest X-rays --- just moderate eclipses. 
In GK Per the pulse profiles are not energy 
dependent, and the pulse fraction is almost the same above and below 10 keV, so we can 
reject this possibility. In GK Per we most probably deal with 
an accretion curtain whose footprint is much extended and forms an arc that covers 180 deg.
The fraction of the arc that is visible can vary smoothly, resulting in only moderate energy
dependence of pulses. It is consistent with the idea proposed by \citet{hel04GKPerRXTE} and 
\citet{vie05GKPer} that in outburst accretion flows to the poles from all azimuths.

Variability on the time-scale of 7000 s was also detected even above 10 keV and 
this cannot be explained by the model of \citet{hel04GKPerRXTE},
in which the QPOs are due to absorption by bulges of material in the accretion disk.
These variations are most probably intrinsic variability of the accretion column emission
due to inhomogeneous accretion. This is supported by the linear dependence between the 
mean X-ray count rate and the pulse amplitude. 

We have also shown that, although the hard X-ray emission shows a very prominent and high-amplitude 
spin modulation, precise measurements of the spin period are not so straightforward due to the 
presence of flickering and the orbital-related variability. 
The {\sl NuSTAR} and Chandra data alone did not allow us to measure the spin period precisely,
and only the long monitoring with {\sl Swift} helped to estimate the spin-up rate. 

\subsection{Hard X-ray component.}
The spectrum above 2 keV can be well fitted with the cooling flow model with maximum 
temperature of 16.2 keV, representing the 
emission from the WD accretion column. 
The continuum indicates that the source is highly absorbed, and this is 
also supported by the fact that the contribution of the cooling flow component to the
observed line emission below 2 keV should be negligible. 
The source of this absorption most probably is the pre-shock material.
Since, depending on the spin phase there will be different amount of absorption in our 
line of sight,
the overall picture is very complex, and is best approximated by the \texttt{pwab} model.

The shock temperature derived from the fit is lower than that 
observed in quiescence and at the beginning of the outburst, which is about 26--27 keV 
\citep{ish92GKPerQui, bru09IPBAT, yua16GKPer}.
When the inner radius of the accretion disc shrinks, the shock temperature is reduced by a 
factor of:
\begin{equation}
f=T_{\rm o}/T_{\rm q}=(1-r_{\rm M o}^{-1})/(1-r_{\rm M q}^{-1})
\end{equation}
\citep{sul05IPmass}, because the approximation of the free fall velocity cannot be used anymore 
(here $r_{\rm M o}$ and $r_{\rm M q}$ are the outburst and quiescent radii of the magnetosphere
in the units of the WD radius). The magnetospheric radius is defined by a balance between
the ram pressure in the disk, which depends on mass transfer, and the magnetic pressure. 
As long as the optical flux is increasing, we expect the mass transfer to be constantly 
increasing as well, since the optical probes the portion of the disc involved in the outburst.
This should result in gradual shrinking of the inner disc radius and lowering of the 
shock temperature. 
The observations in the very first days of the outburst \citep{yua16GKPer} demonstrate
that the decrease of the temperature indeed does not happen immediately.
However, the spectral resolution of the short {\sl Swift} exposures 
and the uncertainty in the intrinsic absorption did not allow us to trace this process.  
Even in the fit to the averaged {\sl Swift} XRT spectra the errors of the plasma
temperature are too high to measure the difference. 


\subsection{Intermediate energy X-ray spectrum}
These are the key points of the analysis of the intermediate X-ray spectral component (0.8--2.0 keV):
\begin{itemize}
\item There are very prominent emission lines of Si, Mg, Ne and \ion{Fe}{xvii}--\ion{Fe}{xviii}.
\item The emission lines did not show any systematic velocity shift or broadening.
\item The emission line ratios do not allow to clearly distinguish between collisional and 
photoionization mechanisms.
\item Assuming collisional ioinzation equilibrium,
the continuum below 2 keV is produced in a medium with a higher temperature than 
the emission line ratios indicate.
\item The spectrum in the 0.8--2 keV range cannot be represented with a model of plasma
in CIE, not with a cooling flow, neither with single or two-temperature \texttt{vapec} model.
\item Neither the continuum nor the emission lines show any spin modulation
 in this energy range.
\end{itemize}

\citet{vie05GKPer} claimed that the emission lines in the X-ray spectrum of GK Per 
originate in the magnetosphere of the WD, because they show no rotation-related
modulations. We found that not only the lines, but also the underlying continuum is not 
modulated, indicating that the source of emission is not confined to the WD polar regions. 
This conclusion is also supported by the estimates of the size of the emitting region, 
which is of the same order of magnitude as the estimated magnetospheric radius of GK Per
by \citet{sul16GKPer} (1.4 R$_{\rm WD}$ and 2.8 R$_{\rm WD}$).
We propose that the magnetospheric boundary is the emission site of the intermediate
spectral component and the intermediate energy X-ray flux is related to the decrease of 
the shock temperature.
If the shock temperature during outbursts is 2/3 of that in quiescence \citep{bru09IPBAT,sul16GKPer}, 
half of the remaining energy is radiated away in the Keplerian disk. Where is the 
remaining 1/6th irradiated and how do we explain the energy budget? The site of emission 
may thus be the magnetospheric boundary, producing this intermediate X-ray spectral component,
in the 0.8--2 keV range, even if the exact mechanism of emission is not clear yet. 
This idea partially explains the anticorrelation of the soft and hard X-rays
during the outburst: the lower the shock temperature, the more energy is released 
in the magnetospheric boundary as moderately soft X-ray emission.
Additionally, \citet{sul16GKPer} measured the magnetospheric radius using the observed 
break frequency in the power spectrum and found $\nu_{\rm break}=0.0225\pm0.004$, corresponding
to the Keplerian velocity of $\sim2500$ km s$^{-1}$, much faster than at the co-rotation 
radius, suggesting that the material in the disc must lose energy in order to decelerate 
and follow the field lines.

\subsection{The soft component.}
The spectral fits and the comparison of the {\sl Chandra} data obtained at different epochs 
(see fig. \ref{fig:comp}) indicate at least two distinct sources of emission below 2 keV. 
However, since there is no proper model for the intermediate energy X-ray spectrum of GK Per, 
it is quite difficult to disentangle the spectral components. The softest part of the spectrum
can indeed be blackbody-like and originate on the surface of the WD, heated by the accretion column. 
Such blackbody-like component in an X-ray spectrum is a distinct property of 
``soft intermediate polars'' \citep[see e.g][]{eva07softIP, anz08softIP}.
The size of the blackbody emitting region was 0.0026 and 0.0083 R$_{\rm WD}$ 
(using the mass-radius relation from \citealt{nau72WDrad} and M$_{\rm}$=0.86 M$_\odot$),
in the first 
and the second halves of the monitoring and the temperature was 60 -- 70 eV, which is within 
the range of typical values for soft IPs. 
The increase of the luminosity of the blackbody-like component indicates that more material
is penetrating deeper in the WD photosphere, producing thermalized X-rays emission,
which may be an effect of the increased mass accretion rate. The latter reaches maximum around
the maximum of the optical light. This is supported by the findings of \citet{sim15GKPerSatur}, 
who analysed a sequence of outbursts in GK Per and noticed a discrepancy between the mass 
transfer through the disc and X-ray emission from the accretion column, which is largest around the 
optical maximum; this can be explained by a buried shock.

On the other hand, there are significant residuals from the blackbody fit of the {\sl Swift} 
XRT spectra (fig. \ref{fig:weeks}). 
Following the discussion from \citet{eva09GKPer} we attempted to estimate the upper limit to the
temperature of the accretion disc in order to check whether it can contribute to the
soft X-ray range. The upper limit to the temperature can be found as:
\begin{equation}
T(R)=\left(\frac{3 GM\dot{M}}{8\pi R^3\sigma}\right)^{1/4}
\end{equation}
\citep{fra02book}. Using the WD mass-radius relation from \citet{nau72WDrad},
the values of $\dot{m}$ from table \ref{tab:nu_spec}, M$_{\rm}$=0.86 M$_\odot$ and inner disc 
radius R$=2.8$R$_{\rm WD}$ \citep{sul16GKPer} we find that the disc temperature can be
as high as 90 000 K. This peak temperature corresponds to a blackbody peak wavelength of 
320 {\AA}. 
The inner disc region significantly contributes to the FUV and UV ranges, 
however, this temperature is still low to be detected with the {\sl Swift} XRT 
(should be at least 150 000 K)
and the disc emission cannot explain the very soft-X-ray excess, seen in fig. \ref{fig:weeks}. 

Strong emission lines were measured in the {\sl Chandra} HETG spectra obtained in 2002 in 
a range as soft as 0.5 keV, where the blackbody component dominates, indicating a significant
contribution from the low-temperature thin plasma emission. The \ion{N}{vii} line at 0.5 keV
additionally shows a completely different profile with two to three components, separated 
by $\sim$1200--1600 km s$^{-1}$.

\section{Conclusions}
We have presented the long-term monitoring of GK Per in a broad energy range,
from UV to hard X-rays, during the dwarf nova outburst in March-April 2015. 
The {\sl NuSTAR} observations allowed to detect a large-amplitude WD spin modulation in 
the very hard X-rays, which is unusual for an IP.

The spectral and timing analysis of our data has revealed 
three distinct spectral components, evolving during the outburst. 
The spectrum above 2 keV can be well explained by a cooling post-shock 
plasma in the accretion column, highly absorbed by a pre-shock material. The spectrum
below $\sim$0.8 keV probably represents the thermalized X-ray emission from the heated WD surface.
 The emission line spectrum 
between 0.8 and 2 keV is the most mysterious, since it cannot be represented by
any existing model of plasma in collisional ionization equilibrium. We propose that 
it originates in the magnetospheric boundary around the WD.

Therefore, as the outburst develops and the mass transfer through the disc grows, 
there are three simultaneous processes affecting the X-ray spectrum:
\begin{itemize}
\item The ram pressure increases at the magnetospheric boundary, pushing the 
accretion disc towards the WD surface and causing the decrease of the shock temperature.
\item The lower the shock temperature with respect to the quiescence level,
the more energy is released in the magnetospheric boundary in the $\sim$0.8 -- 2.0 keV range.
\item Increased specific mass accretion rate in the accretion column results in a higher amount
of material that penetrates deeper in the WD photosphere, causing the increase of the blackbody-like
radiation.
\end{itemize}

The complexity of the X-ray spectrum, the behaviour in different energy ranges and the discrepancy 
between the spectra we obtained and some predictions of the existing models make GK Per a challenging
target for future studies.  
We propose that the observational strategy should be to monitor GK Per at different stages of
its outburst evolution, in order to disentangle the spectral components and to reveal the
contribution from different sources. 

\section*{Acknowledgements}
We acknowledge with thanks the variable star observations from the AAVSO International Data 
base. This work made use of data supplied by the UK Swift Science Data Centre at the University
of Leicester.
Polina Zemko acknowledges a pre-doctoral grant of the CARIPARO 
foundation at the University of Padova.
G. J. M. Luna is a member of the ``Carrera del Investigador Cient\'ifico (CIC)'' of CONICET
and acknowledges support from Argentina grant ANPCYT-PICT 0478/14.
Marina Orio acknowledges an award of the Chandra X-ray Center-Smithsonian Observatory. 
Polina Zemko and Marina Orio also acknowledge support of an INAF-ASI NuSTAR travel grant
 awarded in 2015.


\bsp
\label{lastpage}
\end{document}